\documentclass[12pt]{article}
\usepackage{JASA_manu}
\usepackage[abbr]{jasa_harvard}

\usepackage{amsmath,verbatim,amssymb,epsfig,indentfirst,graphicx,multirow,color,graphics,rotating,url,enumitem}
\usepackage{tabulary}
\definecolor{r}{RGB}{255,0,0}
\definecolor{b}{RGB}{0,0,255}

\begin{document}
\def\T{{ \mathrm{\scriptscriptstyle T} }}
\def\v{{\varepsilon}}
\newcommand{\bph}{\boldsymbol{\Phi}}
\newcommand{\wh}{\widehat}
\newcommand{\mf}{\boldsymbol{\mathrm{f}}}
\newcommand{\mt}{\mathrm{t}}
\newcommand{\me}{\boldsymbol{\mathrm{e}}}
\newcommand{\md}{\boldsymbol{\mathrm{d}}}
\newcommand{\mc}{\boldsymbol{\mathrm{c}}}
\def\ci{\perp\!\!\!\perp}
\def\MGP{multivariate Gaussian processes}
\newcommand{\dmgp}{\mbox{\small dMGP}}
\newcommand{\mgp}{\mbox{\small MGP}}
\newcommand{\bhiw}{\small \mbox{BHIW}}
\newcommand{\iwp}{\mbox{\small IWP}}
\newcommand{\iw}{\mbox{\small IW}}
\newcommand{\hiwp}{\mbox{\small HIWP}}
\newcommand{\tilT}{\widetilde{T}}
\newcommand{\tilcalN}{\widetilde{\mathcal{N}}}
\newcommand{\tilN}{\widetilde{N}}
\newcommand{\calN}{\mathcal{N}}
\newcommand{\normphijk}{||\phi_{jk}||_{L^2(T_j)}^2}
\newcommand{\bUI}{{\bf U}_{I \times I}}
\newcommand{\bs}{\boldsymbol}
\newcommand{\qedwhite}{\hfill \ensuremath{\Box}}
\newcommand{\rev}[1]{\textcolor{blue}{\textsf{#1}}}
\newcommand{\cmt}[1]{\textcolor{red}{\textsf{#1}}}
\newcommand{\bU}{{\bf U}}
\newcommand{\HIWP}{\mbox{\small HIWP}_G}
\newcommand{\shm}{\mbox{\small SHM}}

\baselineskip=25 pt
\title{Bayesian Graphical Models for Multivariate Functional Data}
\author{Hongxiao Zhu$^{1}$, Nate Strawn$^{2}$, and David B. Dunson$^{3}$\\ [.05em]
\small $^1$ Virginia Tech, Blacksburg, VA 24061\\ [.01em]
\small $^2$ Georgetown University, Washington, DC 20057\\ [.01em]
\small $^3$ Duke University, Durham, NC 27708\\ [.01em]}
\date{}
\maketitle


\begin{center}
\bf{Author Footnote:}
\end{center}

Zhu is Assistant Professor, Department of Statistics, Virginia Tech, Blacksburg, VA 24060 (E-mail: {\it hongxiao@vt.edu}). Strawn is Assistant Professor, Department of Mathematics and Statistics, Georgetown University, Washington, DC 20057 (Email: {\it nate.strawn@georgetown.edu}). Dunson is Arts and Sciences Professor, Department of Statistical Science, Duke University, Durham NC 27708 (E-mail: {\it dunson@duke.edu}). This material was based upon work partially supported by the National Science Foundation under Grant DMS-1127914 to the Statistical and Applied Mathematical Sciences Institute. Any opinions, findings, and conclusions or recommendations expressed in this material are those of the author(s) and do not necessarily reflect the views of the National Science Foundation.

\newpage
\begin{center}
\bf{Abstract}
\end{center}

Graphical models express conditional independence relationships among variables.  Although methods for vector-valued data are well established, functional data graphical models remain underdeveloped.  We introduce a notion of conditional independence between random functions, and construct a framework for Bayesian inference of undirected, decomposable graphs in the multivariate functional data context.  This framework is based on extending Markov distributions and hyper Markov laws from random variables to random processes, providing a principled alternative to naive application of multivariate methods to discretized functional data. Markov properties facilitate the composition of likelihoods and priors according to the decomposition of a graph. Our focus is on Gaussian process graphical models using orthogonal basis expansions. We propose a hyper-inverse-Wishart-process prior for the covariance kernels of the infinite coefficient sequences of the basis expansion, establish existence, uniqueness, strong hyper Markov property, and conjugacy. Stochastic search Markov chain Monte Carlo algorithms are developed for posterior inference, assessed through simulations, and applied to a study of brain activity and alcoholism. 



\newpage
\section{Introduction}
\label{sec:intro}
Graphical models provide a powerful tool for describing conditional independence structures between random variables. In the multivariate data case, \citeasnoun{Dawid1993} defined Markov distributions (distributions with Markov property over a graph) of random vectors which can be factorized according to the structure of a graph. They also introduced hyper-Markov laws serving as prior distributions in Bayesian analysis. The special case of Gaussian graphical models, in which a multivariate Gaussian distribution is assumed and 
the graph structure corresponds to the zero pattern of the precision matrix \cite{Dempster1972,Lauritzen1996}, is well studied. Computational algorithms, such as Markov chain Monte Carlo (MCMC) and stochastic search, are developed to estimate the graph based on the conjugate hyper-inverse-Wishart prior and its extensions \cite{Giudici1999,Roverato2002,Jones2005,Scott2008,Carvalho2009}. 

In the frequentist literature, notable works on graphical models include the graphical LASSO \cite{Yuan2007,Friedman2008,mazumder2012a,mazumder2012b} and the neighborhood selection approach \cite{Meinshausen06,ravikumar2010}. The graphical LASSO induces sparse estimation of the precision matrix of the Gaussian likelihood through $l_1$ regularization. The neighborhood selection approach relies on estimating the neighborhood of each node separately by regressing each variable on all the remaining variables, sparsifying with $l_1$ regularization, and then stitching the neighborhoods together to form the global graph estimate. Various extensions, computational methods, and theoretical properties have been developed in these frameworks \cite{lam2009,Hofling2009,Cai2011,Witten2011,Yang2012,mazumder2012a,mazumder2012b,anandkumar2012,loh2013}.

The graphical modeling literature focuses primarily on vector-valued data with each node corresponding to one variable. Many applications, however, involve functional data objects. For example, in neuroimaging, we are often interested in the dependence network across brain regions, where data from each region are of functional form (e.g., EEG/ERP signals, MRI/fMRI regions of interest). 
Although there is an increasingly rich literature on generalizations to accommodate matrix-variate graphical models \cite{Wang2009}, time varying graphical models \cite{Zhou2010,Kolar2011}, and dynamic linear models \cite{Carvalho2007}, the generalization to functional data has not received much attention. 
In recent work, \citeasnoun{Qiao2015} extended graphical LASSO to the functional data case. In this paper, we focus instead on developing Bayesian graphical models for inferring conditional independence structures in multivariate functional data. Most previous work on graphical models has only examined distributions on finite-dimensional metric spaces where many measure-theoretic issues are trivial. Since we must deal with distributions defined on infinite-dimensional spaces, we provide a full measure-theoretic analysis of the constructions and properties.   

In particular, we extend Markov distributions and hyper Markov laws from the random variable to the random process case, facilitating a Bayesian framework for graphical modeling. We then demonstrate the special case of a multivariate Gaussian process in the space of square integrable functions. Through representing the random functions with orthogonal basis expansions, we transform functional data from the function space to the isometrically isomorphic space of basis coefficients, where Markov distributions and hyper Markov laws can be conveniently constructed. We then propose a hyper-inverse-Wishart-process prior for the covariance kernels of the coefficient sequences, and study theoretical properties of the proposed prior, such as existence, uniqueness, the strong hyper Markov property, and conjugacy. To perform posterior inference, we introduce a regularity condition which allows us to write the likelihood and prior density and design stochastic search MCMC algorithms for posterior sampling. Performance of the proposed approach is demonstrated through simulation studies and analysis of brain activity and alcoholism data.    

To our knowledge, the proposed approach is the first  considering functional data graphical models from a Bayesian perspective. It extends the theory of \citeasnoun{Dawid1993} from multivariate data to multivariate functional data. Existing graphical model approaches often naively apply multivariate methods to functional data after performing discretization or feature extraction. Such approaches may not take full advantage of the fact that data arise from a function and can lack reasonable limiting behavior. Our graphical model framework guarantees proper theoretical behavior as well as computational convenience. 


\section{Graphical Models for Multivariate Functional Data} 
\label{sec2:methods}

\subsection{Review of Graph Theory and Gaussian Graphical Models}
\label{sec2:review}
We follow \citeasnoun{Dawid1993}, \citeasnoun{Lauritzen1996}, and \citeasnoun{Jones2005}. Let $G=(V,E)$ denote an undirected graph with a vertex set $V$ and a set of edge pairs $E=\{(i,j)\}$. Each vertex corresponds to one variable. Two variables $a$ and $b$ are conditionally independent if and only if $(a, b) \notin E$.  A graph or a subgraph is {\it complete} if all possible pairs of vertices are joined by edges. A complete subgraph is \emph{maximal} if it is not contained within another complete subgraph. A maximal subgraph is called a {\it clique}. If $A$, $B$, $C$ are subsets of $V$ with $V=A\cup B$, $C=A\cap B$, then $C$ is said to separate $A$ from $B$ if every path from a vertex in $A$ to a vertex in $B$ goes through $C$. $C$ is called a {\it separator} and the pair $(A,B)$ forms a decomposition of $G$.  The separator is {\it minimal} if it does not contain a proper subgraph which also separates $A$ from $B$. While keeping the separators minimal, we can iteratively decompose a graph into a sequence of {\it prime components} -- a sequentially defined collection of subgraphs that cannot be further decomposed \cite{Jones2005}. If all the prime components of a connected graph are complete, the graph is called {\it decomposable}. All the prime components of a decomposable graph are cliques. Iteratively decomposing a decomposable graph $G$ produces a {\it perfectly ordered} sequence of cliques and separators $(C_1,S_2, C_2, \dots, S_m, C_m)$ such that $S_i = H_{i-1}\cap C_i $ and $H_{i-1}=C_1\cup \dots \cup C_{i-1}$. Let $\mathcal{C}=\{C_1,\dots,C_m\}$ denote the set of cliques and $\mathcal{S}=\{S_2,\dots,S_m\}$ denote the set of separators. The perfect ordering means that for every $i=2, \dots, m$, there is a $j<i$ with $S_i\subset C_j$ \cite[page 15]{Lauritzen1996}. 

If the components of a random vector ${\bf X}=(X_1,\dots,X_p)^T$ obey conditional independence according to a decomposable graph $G$, the joint distribution can be factorized as
\[p({\bf X}\mid G) =\frac{\prod_{C\in \mathcal{C}} p({\bf X}_C)}{\prod_{S\in \mathcal{S}} p({\bf X}_S)},\]
where ${\bf X}_A =\{X_i, i\in A\}$. If ${\bf X}$ is Gaussian with zero mean and precision matrix $\boldsymbol{\Omega}=\boldsymbol{\Sigma}^{-1}$, then $X_i$ is conditionally independent of $X_j$ given ${\bf X}_{V \backslash \{i,j\}}$,  denoted by $X_i\ci X_j\mid {\bf X}_{V \backslash \{i,j\}}$, if and only if the $(i,j)$th element of $\boldsymbol{\Omega}$ is zero. In this case $p({\bf X} \mid G)$ is uniquely determined by marginal covariances $\{\boldsymbol{\Sigma}_{C},\boldsymbol{\Sigma}_{S}, C\in \mathcal{C}, S\in \mathcal{S}\}$, which are sub-diagonal blocks of $\boldsymbol{\Sigma}$ according to the clique and separator sets. For a given $G$, a convenient conjugate prior for ${\bf \Sigma}$ is hyper-inverse-Wishart (HIW) with density
\[p(\boldsymbol{\Sigma}\mid G , \delta, {\bf U})=\frac{\prod_{C\in \mathcal{C}} p(\boldsymbol{\Sigma}_C \mid \delta,{\bf U}_C)}{\prod_{S\in \mathcal{S}}p(\boldsymbol{\Sigma}_S\mid \delta,{\bf U}_S)},\]
where $p(\boldsymbol{\Sigma}_C\mid\delta,{\bf U}_C)$ and  $p(\boldsymbol{\Sigma}_S\mid \delta,{\bf U}_S)$ are densities of inverse-Wishart (IW) distributions. In this paper, the inverse-Wishart follows the parameterization of \citeasnoun{Dawid1981}, i.e., $\boldsymbol{\Sigma} \sim \mbox{IW}(\delta, {\bf  U})$ if and only if $\boldsymbol{\Sigma}^{-1}$ has a Wishart distribution $\mbox{W}(\delta+p-1, {\bf U}^{-1})$, where $\delta>0$ and $\boldsymbol{\Sigma}$ is a $p$ by $p$ matrix.

\subsection{Graphical Models for Multivariate Functional Data}
\label{sec2:gpgm_mfd}

Let $\mf=\{ f_j \}_{j=1}^p$ denote a collection of random processes
where each component $f_j$ is in $L^2(T_j)$ and each $T_j$ is a closed subset of the real line. The domain of $\mf$ is denoted by $T = \bigsqcup_{j=1}^p T_j$, where $\bigsqcup$ denotes the disjoint union defined by $\bigsqcup_{j=1}^p T_j = \bigcup_{j=1}^p  \left\{\left(t, j\right): t \in T_j\right\}$. For each $j$, let $\{\phi_{jk}\}_{k=1}^{\infty}$ denote an orthonormal basis of $L^2(T_j)$. 
The extended basis functions $\psi_{jk}=(0,\dots, 0, \phi_{jk},0,\dots,0)$, with $\phi_{jk}$ in the $j$th component and 0 functions elsewhere for $j=1,\dots,p$ and $k=1,\dots, \infty$, form an orthonormal basis of $L^2(T)$.  Let $(L^2(T),\mathcal{B}(L^2(T)), P)$ be a probability space, where $\mathcal{B}(L^2(T))$ is the Borel $\sigma$-algebra on $L^2(T)$.  For $V=\{1, 2, \dots, p\}$ and $A\subset V$, denote by $\mf_A$ the subset of $\mf$ with domain $T_A=\bigsqcup_{j\in A} T_j$. We define the conditional independence relationships for components of $\mf$ in Definition 1. 

{\it Definition 1. }{Let A, B, and C be subsets of V. Then $\mf_A$ is conditionally independent of $\mf_B$ given $\mf_C$ under $P$, written as $\mf_A \ci \mf_B\mid \mf_C [P]$, if for any $\mf_A\in D_A$, where $D_A$ is a measurable set in $L^2(T_A)$, there exists a version of the conditional probability $p(\mf_A\in D_A \mid \mf_B, \mf_C)$ which is $\mathcal{B}(L^2(T_C))$ measurable, and hence one may write $p(\mf_A\in D_A \mid \mf_B,\, \mf_C)= p(\mf_A\in D_A \mid \mf_C)$. Here, $\mathcal{B}(L^2(T_C))$ denotes the Borel $\sigma$-algebra on $L^2(T_C)$. Note that this implies $p(\mf_A\in D_A, \mf_B \in D_B\mid \mf_C)=p(\mf_A\in D_A\mid \mf_C) \, p(\mf_B \in D_B\mid \mf_C)$.}

We would like to use a decomposable graph $G=(V,E)$ to describe the conditional independence relationships of components in $\mf$, whereby a Bayesian framework can be constructed and $G$ can be inferred through posterior inference.  To this end,  we link the probability measure $P$ of $\mf$ with $G$ by assuming that $P$ is {\it Markov} over $G$, as defined in Definition 2.

{\it Definition 2. }{Let $G=(V,E)$ denote a decomposable graph. A probability measure $P$ of $\mf$ is called Markov over $G$ if for any decomposition $(A,B)$ of $G$, $\mf_A \ci \mf_B \mid \mf_{A \cap B}[P]$.}

Given a decomposable graph $G$, a probability measure of $\mf$ with Markov property may be constructed. To enable the construction, we first state Lemma 1, which generalizes Lemma 2.5 of \citeasnoun{Dawid1993} from the random variable to the random process case. 

{\it Lemma 1. }{Let $\mf=(f_1,\dots,f_p)$ be a collection of random processes in $L^2(T)$. For subsets $A,B\subset V=\{1,\ldots,p\}$ with $A\cap B\neq \varnothing$, suppose that $P_1$ and $P_2$ are probability measures of $\mf_A$ and  $\mf_B$, respectively. If $P_1$ and $P_2$ are consistent, meaning that they induce the same measure for $\mf_{A\cap B}$, then there exists a unique probability measure $P$ for $\mf_{A\cup B}$ such that (i) $P_A=P_1$, (ii) $P_B=P_2$, and (iii) $\mf_A\ci \mf_B \mid \mf_{A\cap B}[P]$. The measure $P$ is called a Markov combination of $P_1$ and $P_2$, denoted as $P=P_1\boldsymbol{\star} P_2$. }

With Lemma 1, we can construct a joint probability measure for $\mf$ that is Markov over $G$. The construction is based on the perfectly ordered decomposition $(C_1,S_2, C_2,$ $\ldots, S_m, C_m)$ of $G$ with $S_i = H_{i-1}\cap C_i $ and $H_{i-1}=C_1\cup \dots \cup C_{i-1}$. Let $\{M_{C_i}, i=1,\dots, m\}$ be a sequence of pairwise consistent probability measures for $\{\mf_{C_i}, i=1,\dots, m\}$. We construct a Markov probability measure $P$ over $G$ through the following recursive procedure
\begin{eqnarray}
P_{C_1} & = & M_{C_1}, \label{eq1}\\ 
P_{H_{i+1}} & = & P_{H_i}\boldsymbol{\star} M_{C_{i+1}},\quad i=1,\dots, m-1.
\label{eq2}
\end{eqnarray}
One can show that the probability measure constructed this way is the unique Markov distribution over $G$ with marginals $\{M_{C_i}\}$, and the proof follows that of Theorem 2.6 in \citeasnoun{Dawid1993}. We call the constructed probability measure the {\it Markov distribution} of $\mf$ over $G$.

Denote the Markov distribution of $\mf$ constructed in (\ref{eq1}) - (\ref{eq2}) by $P_G$, and denote the space of all Markov distributions over $G$ by $\mathcal{M}(G)$.  A prior law for $P_G$ is then supported on $\mathcal{M}(G)$. We follow \citeasnoun{Dawid1993} to define hyper Markov laws and use them as prior laws for $P_G$. A prior law $\mathfrak{L}$ of $P_G$ is called {\it hyper Markov} over $G$ if for any decomposition $(A,B)$ of $G$, $(P_G)_A \ci (P_G)_B \mid (P_G)_{A\cap B}[\mathfrak{L}]$, where $(P_G)_A$ takes values in $\mathcal{M}(G_A)$ which is the space of all Markov distributions over subgraph $G_A$. Here, we have assumed that $G$ is collapsible onto A, therefore $\phi \in \mathcal{M}(G_A)$ if and only if $\phi = (P_G)_A$ for some $(P_G) \in \mathcal{M}(G)$. 
The following Proposition 1 states that the theory of hyper Markov laws of \citeasnoun{Dawid1993} applies to our random process setup. 

{\it Proposition 1. }{The theory of hyper Markov laws over undirected decomposable graphs, as described in Section 3 of \citeasnoun{Dawid1993}, holds for random processes.}

According to the theory of hyper Markov laws, one can construct a prior law for $P_G$ using a sequence of consistent marginal laws $\{\mathfrak{L}_C, C\in \mathcal{C}\}$ in a similar fashion as (\ref{eq1}) - (\ref{eq2}). Denote by $\mathfrak{L}_G$ the constructed hyper Markov prior for $P_G$ and by $\Pi$ a prior distribution for the graph $G$. A Bayesian graphical model for the collection of random processes $\mf$ can be described as 
\begin{eqnarray}
\mf  \sim P_G; \quad P_G  \sim \mathfrak{L}_G; \quad G \sim  \Pi.
\label{eq:bayes_frame}
\end{eqnarray}
As we have yet to specify a concrete example for the probability measure $P_G$, the above Bayesian framework remains abstract at the moment. In Section \ref{sec:gprocess_mfd}, we construct $P_G$ using Gaussian processes and propose a hyper-inverse-Wishart-process law as the prior for $P_G$. The prior distribution $\Pi$ is supported on the finite dimensional space of decomposable graphs with $p$ nodes.

\subsection{Gaussian Process Graphical Models for Multivariate Functional Data}
\label{sec:gprocess_mfd}
Let $\mf_0=(f_{01},\dots,f_{0p})$ be an element in $L^2(T)$. Denote by $\mathcal{K}=\{k_{ij}:T_i\times T_j \rightarrow \mathbb{R}\}$ a collection of covariance kernels such that $\mbox{cov}\{f_i(s), f_j(t)\}=k_{ij}(s,t),  s\in T_i, t\in T_j$. 
We assume that $\mathcal{K}$ is positive semidefinite and trace class. Positive semidefinite means that
\[
\sum_{i,j=1}^p\sum_{k,l=1}^\infty c_{ik}c_{jl}\int_{T_j}\int_{T_i}k_{ij}(s,t)\phi_{ik}(s)\phi_{jl}(t)dsdt\geq0
\]
for any square summable sequence $\{c_{ik},\, i=1,\dots,p, \, k=1,\dots, \infty\}$; trace class means that
\[
\sum_{j=1}^p\sum_{l=1}^\infty\int_{T_j}\int_{T_i}k_{jj}(s,t)\phi_{jl}(s)\phi_{jl}(t)dsdt<\infty.
\]
Then $\mf_0$ and $\mathcal{K}$ uniquely determine a Gaussian process on $L^2(T)$ \cite{Prato2006}, which we call a multivariate Gaussian process, and write $\mgp( \mf_0, \mathcal{K} )$. The definition of multivariate Gaussian process implies that for $A\subset V$, $\mf_A \sim \mgp(\mf_{0A},\mathcal{K}_A)$ where $\mathcal{K}_A=\{k_{ij}, i,j\in A\}$. Furthermore, on a sequence of cliques $\mathcal{C}=\{C_1,\dots, C_m\}$, the marginal Gaussian process measures for $\{\mf_{C}, C\in \mathcal{C}\}$ are automatically consistent because they are induced from the same joint distribution. Therefore, we can construct a Markov distribution for $\mf$ over $G$ through procedure (\ref{eq1}) - (\ref{eq2}). We denote the resulting 
distribution of $\mf$ by $\mgp_G( \mf_0, \mathcal{K}_{\mathcal{C}} )$, where $\mathcal{K}_{\mathcal{C}} = \{k_{ij}: i,j\in C, C\in \mathcal{C}\}$. It is clear from this construction that the distribution $\mgp_G$ is Markov over $G$ whereas $\mgp$ is not. 

For the convenience of both theoretical analysis and computation, we represent elements in $L^2(T)$ using orthonormal basis expansions and construct a Bayesian graphical model in the dual space of basis coefficients. Let $\{\phi_{jk}\}_{k=1}^\infty$ denote an orthonormal basis of $L^2(T_j)$, and $f_j(t)=\sum_{k=1}^\infty c_{jk}\phi_{jk}(t)$ where $c_{jk}=\langle f_j, \phi_{jk} \rangle =\int_{T_j} f_j(t)\phi_{jk}(t)dt$. The coefficient sequence $c_j=\{c_{jk}, k=1,\dots,\infty\}$ lies in the space of square-summable sequences, denoted by $\ell_j^2=\left\{c_{jk}:\sum_{k=1}^\infty c_{jk}^2<\infty\right\}$. Denote $\ell^2= \prod_{j=1}^p \ell_j^2$. Since $\ell_j^2$ and $L^2(T_j)$ are isometrically isomorphic for each $j$, once an orthonormal basis of $L^2(T)$ has been chosen, we have an identification between the Borel probability measures defined on $\ell^2$ and $L^2(T)$; therefore we can construct statistical models on $\ell^2$ without loss of generality. 
Let $\mc=(c_1,\dots,c_p)$ denote the coefficient sequence of $\mf$. Then $\mf \sim \mgp (\mf_0, \mathcal{K})$ corresponds to $\mc \sim \dmgp(\mc_0, \mathcal{Q})$, where $\dmgp$ denotes the infinite dimensional discrete \MGP, $\mc_0$ is the coefficient sequence of $\mf_0$ and $\mathcal{Q}=\{q_{ij}(\cdot,\cdot), i,j \in V\}$. Here, $q_{ij}$ is the covariance kernel so that $\mbox{cov}(c_{ik},c_{jl})=q_{ij}(k,l)$ for $k, l\in \{1,2,3, \ldots\}$. Similarly, $\mf \sim \mgp_G( \mf_0, \mathcal{K}_{\mathcal{C}} )$ corresponds to $\mc \sim \dmgp_G(\mc_0, \mathcal{Q}_{\mathcal{C}})$ where $\mathcal{Q}_{\mathcal{C}}=\{q_{ij}(\cdot,\cdot), i,j \in C, C\in \mathcal{C}\}$. The collection $\mathcal{Q}$ is also positive semidefinite and trace class, so that $
\sum_{i,j=1}^p\sum_{k,l=1}^\infty c_{ik}c_{jl} q_{ij}(k,l)\geq0$ for any square summable sequence $\{c_{ik}, i=1,\dots,p, k=1,\dots, \infty\}$, and $\sum_{j=1}^p \sum_{k=1}^\infty q_{jj}(k,k)<\infty$. Furthermore,  $\mathcal{K}$ relates to $\mathcal{Q}$ through equation $k_{ij}(s,t)=\sum_{k,l=1}^{\infty}q_{ij}(k,l)\phi_{ik}(s)\phi_{jl}(t)$. Denote by $P^{\mc}$ and $P^{\mf}$ the probability measures of $\mc$ and $\mf$ respectively, then $\mf_A \ci \mf_B\mid \mf_C [P^{\mf}]$ implies $\mc_A \ci \mc_B\mid \mc_C [P^{\mc}]$ and vice versa. Thus, the distribution $ \dmgp_G(\mc_0, \mathcal{Q}_{\mathcal{C}})$ of $\mc$ is again Markov. 

Assume that $\mc \sim \dmgp_G(\mc_0, \mathcal{Q}_{\mathcal{C}})$. 
The parameters involved in this distribution include $\mc_0$ and $\mathcal{Q}_{\mathcal{C}}$. In this study, we assume that $\mc_0$ is fixed (e.g., a zero sequence) so that the distribution of $\mc$ is uniquely determined by $\mathcal{Q}_{\mathcal{C}}$. As indicated in Section \ref{sec2:gpgm_mfd}, we would like to construct a hyper Markov law for the $\dmgp_G$ distribution. Since $\dmgp_G$ is uniquely determined by $\mathcal{Q}_{\mathcal{C}}$, it is equivalent to construct a hyper Markov law for $\mathcal{Q}_{\mathcal{C}}$. Given a positive integer $\delta$ and a collection $\mathcal{U}=\{u_{ij}: \mathbb{N}\times \mathbb{N} \rightarrow \mathbb{R}, i, j \in V\}$ which is symmetric, positive semidefinite, and trace class, we construct a hyper-inverse-Wishart-process (HIWP) prior for $\mathcal{Q}_{\mathcal{C}}$ following Theorem 1.  

{\it Theorem  1. }{Assume that $\mc \sim \dmgp_G(\mc_0, \mathcal{Q}_{\mathcal{C}})$. Suppose that $\delta$ is a positive integer, and $\mathcal{U}$ is a collection of kernels that is symmetric, positive semidefinite and trace class. Then there exists a sequence of pairwise consistent inverse-Wishart processes determined by $\delta$ and $\mathcal{U}_C=\{u_{ij}, i,j\in C\}, C\in \mathcal{C}$, based on which one can construct a unique hyper Markov law for $\mathcal{Q}_{\mathcal{C}}$, which we call a hyper-inverse-Wishart-process, and write $\mathcal{Q}_{\mathcal{C}} \sim \mbox{\small HIWP}_G(\delta, \mathcal{U}_{\mathcal{C}})$, where $\mathcal{U}_{\mathcal{C}}=\{u_{ij}, i,j \in C, C \in \mathcal{C}\}$.}    

Based on Theorem 1, a Bayesian Gaussian process graphical model can be written as 
\begin{eqnarray}
\mc \sim \dmgp_G(\mc_0,\mathcal{Q}_{\mathcal{C}}), \quad \mathcal{Q}_{\mathcal{C}}\sim \mbox{\small HIWP}_G(\delta,\mathcal{U}_{\mathcal{C}}), \quad
G \sim \Pi.
\label{eq:mod_1}
\end{eqnarray}
It is of interest to investigate the properties of the HIWP prior and the corresponding posterior distribution. As shown in \citeasnoun{Dawid1993}, one nice property of the HIW law is the strong hyper Markov property, which leads to conjugacy as well as convenient posterior computation at each clique. In case of the HIWP prior, the strong hyper Markov property is defined such that for any decomposition $(A, B)$ of $G$ in model (\ref{eq:mod_1}), $\mathcal{Q}_{B | A} \ci \mathcal{Q}_A$, where $\mathcal{Q}_{B|A}$ denotes the conditional distribution (i.e., conditional covariance) of $\mc_B$ given $\mc_A$. In the following proposition, we show that the $\mbox{\small HIWP}_G$ prior constructed in Theorem 1 is strong hyper Markov 
when $\mbox{rank}(u_{ij})<\infty$ for $i, j \in V$.

{\it Proposition 2. }{Suppose that the collection of kernels $\mathcal{U}$
satisfies that \mbox{rank}$(u_{ij}) <\infty$ for $i, j \in V$, then the hyper-inverse-Wishart-process prior constructed in Theorem 1 satisfies the strong hyper Markov property. That is, if $\mathcal{Q}_{\mathcal{C}} \sim \mbox{\small HIWP}_G(\delta, \mathcal{U}_{\mathcal{C}})$, then for any decomposition $(A, B)$ of $G$, $\mathcal{Q}_{B | A} \ci \mathcal{Q}_A$, where $\mathcal{Q}_{B|A}$ denotes the conditional distribution (e.g., conditional covariance) of $\mc_B$ given $\mc_A$.} 
  
The strong hyper Markov property of $\mbox{\small HIWP}_G$ ensures that the joint posterior of $\mathcal{Q}_{\mathcal{C}}$ (conditional on $G$) can be constructed from the marginal posterior of $\mathcal{Q}_{C}$ (conditional on $G$) at each clique $C$, as stated in Theorem 2. Therefore one essentially transforms the Bayesian analysis to a sequence of sub-analyses at the cliques, which substantially reduces the size of the problem.

{\it Theorem 2. }{Suppose that $\mc_i\sim \dmgp_G(\mc_0,\mathcal{Q}_{\mathcal{C}}), i=1,\ldots,n$ are independent and identically distributed. Further assume that the prior of $\mathcal{Q}_{\mathcal{C}}$ is $\mbox{\small HIWP}_G(\delta,\mathcal{U}_{\mathcal{C}})$ where the collection of kernels $\mathcal{U}$ satisfies that \mbox{rank}$(u_{ij}) <\infty$ for $i, j \in V$. Then the conditional posterior of $\mathcal{Q}_{\mathcal{C}}$ given $\{{\mc}_i\}$ and $G$ is $\mbox{\small HIWP}_G(\widetilde{\delta},\widetilde{\mathcal{U}}_{\mathcal{C}})$, where $\widetilde{\delta}=\delta+n$, $\widetilde{\mathcal{U}}_{\mathcal{C}}=\{\widetilde{u}_{ij}, i,j \in C, C \in \mathcal{C}\}$ and   $\widetilde{u}_{ij}={u}_{ij}+\sum_{i=1}^n (\mc_i-\mc_{0i}) \otimes({\mc}_j-{\mc}_{0j})$. Here $\otimes$ denotes the outer product.  Furthermore, the marginal distribution of $\{{\mc}_i\}$ given $\{G, \mc_0, \delta,  \, \widetilde{\mathcal{U}}_{\mathcal{C}}\}$ is again Markov over $G$.}

Theorem 2 implies that when \mbox{rank}$(u_{ij}) <\infty$ for $i, j \in V$, the $\mbox{\small HIWP}_G(\delta, \mathcal{U}_{\mathcal{C}})$ prior is a conjugate prior for $\mathcal{Q}_{\mathcal{C}}$ in the $\dmgp_G(\mc_0,\mathcal{Q}_{\mathcal{C}})$ likelihood. Note that here the likelihood, the prior, and the posterior are all conditional on $G$, which makes Bayesian inference of $G$ tractable. Model (\ref{eq:mod_1}) and results in Theorem 2 provide the theoretical foundation for practical Bayesian inference under a reasonable regularity condition, as discussed in Section \ref{sec:approx_infer}.

\section{Bayesian posterior inference}
\label{sec:approx_infer}

Despite the fact that functional data are realizations of inherently infinite-dimensional random processes,  data can only be collected at a finite number of measurement points. Essentially, estimating the conditional independence structure of infinite-dimensional random processes based on a finite number of measurement points is an inverse problem and therefore requires regularization.  \citeasnoun{Muller2008} reviewed two main approaches for regularization in functional data analysis---finite approximation through, e.g., suitably truncating the basis expansion representation and penalized likelihood. In this paper, we suggest performing posterior inference based on approximating the underlying random processes with orthogonal basis functions. In particular, we assume the following regularity condition:   

{\it Condition 1. }{The functional data $\mf$ are observed discretely on a 
dense grid ${\bf t}= \bigsqcup{\bf t}_j$ with ${\bf t}_j=(t_{j1},\dots, t_{jm_j(n)})$ and $m_j(n)\rightarrow \infty$ as $n\rightarrow \infty$. One can find $M_j (n)$ so that the underlying random process $f_j$ can be approximated with an $M_j$-term orthogonal basis expansion $\widehat{f}_j =\sum_{l=1}^{M_j} c_{jl}\phi_{jl}$, with approximation error $||f_j-\widehat{f}_j||_{L^2} = O_p (n^{-\beta})$ with $\beta\geq 1/2$ for all $j\in V$.}

Essentially, Condition 1 requires that the discretely-measured functional data capture sufficient information about the underlying random processes, so that we can approximate each $f_j$ with a negligible approximation error. Condition 1 is a basic assumption in the functional setting, and a similar regularity condition has been adopted by \citeasnoun{Qiao2015} in a functional graphical model based on the group LASSO penalty.

\subsection{Bayesian Posterior Inference under the Regularization Condition}
\label{approx:inf}

The regularity from Condition 1 enables us to write the density functions of the Markov distributions and hyper Markov laws so that posterior inference can be practically implemented. Denoting $M=(M_1,\dots,M_p)$, we can explicitly write the density function for the truncated process $\mc^M = (c_1^{M_1},\ldots,c_p^{M_p})$, and an MCMC algorithm can then be designed for the posterior inference of the underlying graph $G$. The density function of $\mc^M$ is
\begin{eqnarray}
p(\mc^M \mid {\mc}_0^M, \textbf{Q}_{\mathcal{C}}, G)=\frac{\prod_{C\in\mathcal{C}}p(\textbf{c}_C^M \mid \textbf{c}_{0,C}^M, \textbf{Q}_C)}{\prod_{S\in\mathcal{S}}p(\textbf{c}_S^M\mid \textbf{c}_{0,S}^M, \textbf{Q}_S)},\label{eq:density}
\end{eqnarray}   
where $\textbf{Q}_{\mathcal{C}}$ is a block-wise covariance matrix with the $(i,j)$th block formed by $\{q_{ij}(k,l), k=1,\dots, M_i, l=1,\dots, M_j\}$, and $\textbf{Q}_C$, $\textbf{Q}_S$ are submatrices of $\textbf{Q}_{\mathcal{C}}$ corresponding to clique $C$ and separator $S$, respectively. 
The $\mbox{\small HIWP}_G$ prior of $\mathcal{Q}_{\mathcal{C}}$ induces a hyper inverse-Wishart prior with density
\begin{eqnarray}
p(\textbf{Q}_{\mathcal{C}}\mid G)=\frac{\prod_{C\in \mathcal{C}} p( \textbf{Q}_C \mid \delta, \textbf{U}_C )}{\prod_{S\in \mathcal{S}} p(\textbf{Q}_S \mid \delta, \textbf{U}_S)},
\label{prior:hiw0} 
\end{eqnarray}
where $p(\textbf{Q}_C \mid \delta, \textbf{U}_C )$ is the density of inverse-Wishart defined in \citeasnoun{Dawid1981}, $\textbf{U}_C$ is a submatrix of $\textbf{U}_{\mathcal{C}}$ corresponding to clique $C$, and $\textbf{U}_{\mathcal{C}}$ is a block-wise matrix formed by $\{u_{ij}\}$ in the same way as $\textbf{Q}_{\mathcal{C}}$ is formed by $\{q_{ij}\}$. The $p(\textbf{Q}_S \mid \delta, \textbf{U}_S)$ component in the denominator  is defined similarly. Based on (\ref{eq:density}) and (\ref{prior:hiw0}), and assuming that $\{\textbf{c}_i, i=1,\dots, N\}$ is a random sample of $\textbf{c}$,  one can further integrate out $\textbf{Q}_{\mathcal{C}}$ to get the marginal density
\begin{eqnarray}
p(\{\textbf{c}_i^M\}\mid \textbf{c}_0^M,G)
=(2\pi)^{-\frac{N}{2}(\sum_i M_i)} \frac{h(\delta,\textbf{U}_{\mathcal{C}})}{h(\widetilde{\delta}, \widetilde{\textbf{U}}_{\mathcal{C}})},
\label{eq:intsig}
\end{eqnarray}
where
\[h(\delta,\textbf{U}_{\mathcal{C}})=\frac{\prod_{C\in\mathcal{C}}|\frac{1}{2}\textbf{U}_C|^{(\frac{\delta+d_c-1}{2})}\Gamma_{d_c}^{-1}\{\frac{1}{2}(\delta+d_c-1)\}}{\prod_{S\in \mathcal{S}}|\frac{1}{2} \textbf{ U}_S|^{(\frac{\delta+d_s-1}{2})}\Gamma_{d_s}^{-1}\{\frac{1}{2}(\delta+d_s-1)\}},\]
and $d_c$ and $d_s$ are the dimensions of $\textbf{U}_C$ and $\textbf{U}_S$ respectively, and $\Gamma_b(a)=\pi^{b(b-1)/4}\prod_{i=0}^{b-1}\Gamma(a-i/2)$. The denominator $h(\widetilde{\delta}, \widetilde{\textbf{U}}_{\mathcal{C}})$ in (\ref{eq:intsig}) is defined in the same way. Based on these results, posterior inference can be done through sampling from the posterior density
\begin{eqnarray}
p(G\mid \{\textbf{c}_i^M\},\textbf{c}_0^M) \propto p(\{\textbf{c}_i^M\}\mid \textbf{c}_0^M, G)\, p(G),
\label{eq:mod2}
\end{eqnarray}
where $p(G)$ is the density function corresponding to the prior distribution $G \sim \Pi$, which is a discrete distribution supported on all decomposable graphs with $p$ nodes.  \citeasnoun{Giudici1999} used the discrete uniform prior $\mbox{Pr}(G=G_0)=1/d$ for any fixed $p$-node decomposable graph $G_0$, where $d$ is the total number of such graphs; \citeasnoun{Jones2005} used the independent Bernoulli prior with probability $2/(p-1)$ for each edge, which favors sparser graphs \cite{Giudici1996}. The following MCMC algorithm describes the steps to generate posterior samples based on (\ref{eq:mod2}).

\medskip
\noindent
\hspace{-0.8 mm}{\it Algorithm 1.}
\begin{enumerate}[leftmargin=3.4 pc]
\item[\it Step 0.] Set an initial decomposable graph $G$ and set the prior parameters $\textbf{c}_0$, $\delta$, and $\textbf{U}_{\mathcal{C}}$.
\item[\it Step 1.] With probability $1-q$, propose $\widetilde{G}$ by randomly adding or deleting an edge from $G$ (each with probability $0.5$) within the space of decomposable graphs; with probability $q$, propose $\widetilde{G}$ from a discrete uniform distribution supported on the set of all decomposable graphs. Accept the new $\widetilde{G}$ with probability
\[\alpha=\min\left \{1, \frac{p(\widetilde{G}\mid \{\textbf{c}_i^M\},\textbf{c}_0^M)\,\, p(G\mid \widetilde{G})}{p({G}\mid \{\textbf{c}_i^M\},\textbf{c}_0^M)\, \,p(\widetilde{G}\mid G)}\right\}.\]
\end{enumerate}
Repeat step 1 for a large number of iterations until convergence is achieved.

Detailed derivations are available in the Supplementary Materials. The above algorithm is a Metropolis-Hastings sampler with a mixture of local and heavier-tailed proposals, also called a {\it small-world sampler}. The ``local" move involves randomly adding or deleting one edge based on the current graph, and the ``global" move is achieved through the discrete uniform proposal. \citeasnoun{Guan2006} and  \citeasnoun{Guan2007} have shown that the small-world sampler leads to much faster convergence especially when the posterior distribution is either multi-modal or spiky.

\subsection{Bayesian Posterior Inference for Noisy Functional Data}
\label{sec2:meserror}

The theory in Section \ref{sec2:methods} and the posterior inference in Section \ref{approx:inf} relies on the assumption that the distribution of $\mf$ (and $\mc$) is Markov over $G$. In many situations, it is more desirable to make such an assumption in a hierarchical model. For example, when functional data are subject to measurement error, one might wish to incorporate an additive error term and consider the following model for the coefficient process:
\begin{equation}
\label{eq:base}
d_{ijk} = c_{ijk}+e_{ijk}, \quad i=1,\dots,N, \quad j=1\dots,p, \quad k=1,\dots, \infty,
\end{equation}
where $\{c_{ijk}\}$ and $\{e_{ijk}\}$ are mutually independent with Gaussian distributions. This induces an additive model in the $L^2(T)$ space:  $y_{ij}=f_{ij}+\v_{ij}$, where $\{y_{ij}\}$ are the functional data observations, $\{f_{ij}\}$ are the underlying true functions and $\{\v_{ij}\}$ are residuals. We assume $e_{ijk} \sim N(0, s_j^2)$ which corresponds to assuming white noise for $\v_{ij}$. After concatenating the $p$ coefficient sequences to vector forms, we obtain the model 
$\md_i=\mc_i+\me_i$, where $\md_i=(\mathrm{d}_{i1},\dots,\mathrm{d}_{ip})$, $\mathrm{d}_{ij}=(d_{ij1},d_{ij2},\dots)$, and $\mc_i$, $\me_i$ follow similar forms. 

After truncation at $M$, $\me_i^M \sim N(0, \boldsymbol{\Lambda})$ and $\boldsymbol{\Lambda}=\mbox{diag}(s_1^2 1_{M_1}^T,\dots,s_p^2 1_{M_p}^T)$.
Notice that here $\mbox{cov}({\md}_i^M)={\bf Q}_{\mathcal{C}}+\boldsymbol{\Lambda}$, thus the diagonals of ${\bf Q}_{\mathcal{C}}$ and $\boldsymbol{\Lambda}$ can not be separately identifiable. Therefore, we treat $\boldsymbol{\Lambda}$ as a fixed model parameter, whose quantity can be pre-determined through the approximation: $s_j^2\approx \widehat{\sigma}_j^2|T_j|/(|\boldsymbol{\mt}_j|-1)$, where $\widehat{\sigma}_j^2$ is the estimated variance of $\v_{ij}$ using local smoothing, $|T_j|$ is the width of interval $T_j$, and $|\boldsymbol{\mt}_j|$ is the number of grid points in $T_j$.
Applying a prior for $\mc_i^M$ in the form of (\ref{eq:density}) (conditional on $G$) and the $\mbox{\small HIWP}_G$ prior for the covariance matrix $\mathcal{Q}_{\mathcal{C}}$ in the form of (\ref{prior:hiw0}), we obtain the density function for the joint posterior:  
\begin{eqnarray}
 p(\{\mc_i^M\},{\bf Q}_{\mathcal{C}},G \mid \{\md_i^M\}) 
 \propto \prod_{i=1}^n p({\md}_i^M\mid {\mc}_i^M, \boldsymbol{\Lambda})\; p({\mc}_i^M\mid {\mc}_0^M,{\bf Q}_{\mathcal{C}}, G)\;p({\bf Q}_{\mathcal{C}}\mid G)\; p(G).  
\label{postexpre}
\end{eqnarray}From (\ref{postexpre}), we can integrate out ${\bf Q}_{\mathcal{C}}$ to obtain the marginal posterior distribution of $\{{\mc}_i^M\}$ and $G$. The MCMC algorithm for generating posterior samples based on (\ref{postexpre}) is listed in Algorithm 2.

\medskip
\noindent
\hspace{-0.8 mm}{\it Algorithm 2.}
\begin{itemize}[leftmargin=3.4 pc]
\item[\it Step 0] Set initial values for $\{{\mc}_i^M\}$, $G$ and set the model parameters $\delta$,  ${\mc}_0^M$, ${\bf U}$ and $\boldsymbol{\Lambda}$.
\item[\it Step 1] Conditional on $\{{\mc}_i^M\}$, update $G\sim p(G\mid\{{\mc}_i^M\}, \mc_0^M )$ using the small-world sampler as described in Step 1 of Algorithm 1, where $p(G\mid\{{\mc}_i^M\},\mc_0^M)$ is computed based on (\ref{postexpre}).
\item[\it Step 2] Given $G$, update ${\bf Q}_{\mathcal{C}} \sim p({\bf Q}_{\mathcal{C}} \mid\{{\mc}_i^M\},G)$, which takes the same form as (\ref{prior:hiw0}) except that $\delta$ and ${\bf U}$ are replaced by $\widetilde{\delta}$ and $\widetilde{\bf U}$ respectively using the formulae in Theorem 2. 
\item[\it Step 3] Conditional on $G$ and ${\bf Q}_{\mathcal{C}}$, update ${\mc}_i^M \sim N(\boldsymbol{\mu}_i, {\bf V})$, where ${\bf V}=(\boldsymbol{\Lambda}^{-1}+{\bf Q}_{\mathcal{C}}^{-1})^{-1}$ and $\boldsymbol{\mu}_i={\bf V}(\boldsymbol{\Lambda}^{-1}{\md}_i^M + {\bf Q}_{\mathcal{C}}^{-1}{\mc}_0^M)$.
\end{itemize}
Repeat step 1--3 for a large number of iterations until convergence is achieved.

\subsection{Other Practical Computational Issues}
\label{sec3:fit}

Calculating the coefficient sequences $\{\mc_i\}$ from the functional observations $\{\mf_i\}$ requires the selection of an orthonormal basis $\{\phi_{jk}, j=1,\dots, p, k=1,\dots,\infty\}$. If a known basis is chosen (e.g., Fourier), the coefficient sequences can be estimated by $c_{ijk}=\langle f_{ij}, \phi_{jk}\rangle$ using numerical integration. Another convenient choice is the eigenbasis of the autocovariance operators of $\{\mf_i\}$, in which case the coefficient sequences are called functional principal component (FPC) scores. The corresponding basis representation is called Karhunen-Lo\`{e}ve expansion. The eigenbasis can be estimated using the method of \citeasnoun{Ramsay2005} or the Principal Analysis by Conditional Expectation (PACE) algorithm of \citeasnoun{Yao2005}. Owing to the rapid decay of the eigenvalues, the eigenbasis provides a more parsimonious and efficient representation compared with other bases. Furthermore, the FPC scores within a curve are mutually uncorrelated, so one may set the prior parameter ${\bf U}_{\mathcal{C}}$ to be a matrix with blocks of diagonal sub-matrices, or simply a diagonal matrix. 

In addition to the estimation of coefficient sequences, a suitable truncation of the infinite sequences $\{\mc_i\}$ is needed to facilitate practical  posterior inference. We suggest to pre-determine the truncation parameters using approximation criteria, following \citeasnoun{RiceSilverman1991}, \citeasnoun{Yao2005}, or \citeasnoun{Li2013}. This includes cross-validation \cite{RiceSilverman1991}, applying the Akaike information criterion or Bayesian information criterion \cite{Yao2005,Li2013}, or controlling the fraction-of-variance-explained (FVE) in the FPC analysis \cite{Lei2014}.

\section{Simulation Study}
\label{sec4:simu}

Three simulation studies were conducted to assess the performance of posterior inference using the Gaussian process graphical models outlined in Section \ref{sec:gprocess_mfd} and Section \ref{sec:approx_infer}. Simulation 1 corresponds to the smooth functional data case (without measurement error), and Simulation 2 corresponds to the noisy data case when measurement error is considered. Both simulations are based on a true underlying graph with $6$ nodes, demonstrated in Figure~\ref{fig:simudata} (a). In simulation 3, we show the performance of the proposed Bayesian inference in a $p>n$ case, with the number of nodes $p=60$ and the sample size $n = 50$.

\subsection{Simulation 1: Graph Estimation for Smooth Functional Data}
\label{sec:simu1}
\begin{figure}
\centering
\includegraphics[width=6 in, angle=0]{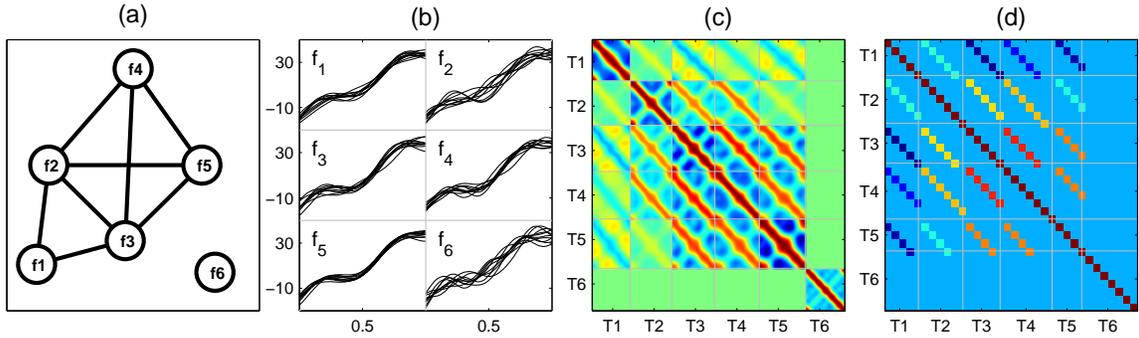}
\caption[Plots of Simulation 1]{Plots of Simulation 1: (a) The true underlying graph; (b) The first $10$ samples of $\{f_{ij}, j=1,\dots, 6\}$; (c) The image plot of the underlying data-domain correlation matrix; (d) The image plot of the underlying correlation matrix ${\bf R}$.}
\label{fig:simudata}
\end{figure}

Multivariate functional data are generated on the domain $[0,1]$ using Fourier basis with the number of basis functions $\{M_j\}_{j=1}^p$ varying from $3$ to $7$. The true eigenvalues are generated from Gamma distributions and are subject to exponential decay. The conditional independence structure is determined by a $p\times p$ correlation matrix ${\bf R}_0$, with the inverse ${\bf R}_0^{-1}$ containing a zero pattern corresponding to the graph in Figure~\ref{fig:simudata} (a). We then generate principal component scores from a multivariate normal distribution with zero mean and a block-wise covariance matrix ${\bf Q}={\bf Z}{\bf R} {\bf Z}$, which has dimension $\sum_{j=1}^p M_j$. Here ${\bf R}$ is a block-wise correlation matrix that has a diagonal form in each block. In particular, the $(i,j)th$ block of ${\bf R}$, denoted by ${\bf R}_{ij}$, satisfies 
that ${\bf R}_{ij}=({\bf R}_0)_{i,j}{\bf I}$ where ${\bf I}$ is a rectangular identity matrix with size $M_i \times M_j$. An image plot of ${\bf R}$ is shown in Figure~\ref{fig:simudata}(d), with its data-domain counterpart (the correlation of $\mf$ evaluated on a grid ${\bf t}$) shown in Figure~\ref{fig:simudata}(c). The multivariate functional data are finally generated through linearly combining the eigenbasis using the principal component scores. A common mean function is added to each curve.  The generated data contain $n=200$ independent samples, and each sample contains six curves measured on six different grids. We display the first $10$ samples in Figure~\ref{fig:simudata}(b).

Based on the data generated above, we estimate the principal component scores $\{\mc_i\}$ using the PACE algorithm of \citeasnoun{Yao2005} and determine the truncation parameter $\{M_j\}$ using the FVE criterion with a $90\%$ threshold, resulting in $\{M_j\}$ values around $5$. We apply Algorithm 1 and set $\delta=5$ and ${\bf U}={\widehat{\bf Z}\widehat{\bf R}\widehat{\bf Z}}$, where ${\widehat{\bf Z}}=\mbox{diag}\{\widehat{\lambda}_{jk}^{1/2},k=1,\dots,M_j, j=1,\dots, p\}$, $\{\widehat{\lambda}_{jk}\}$ are the estimated eigenvalues and $\widehat{\bf R}$ is set to be the identity marix.  A total of $5,000$ MCMC iterations are performed. Starting from the empty graph, the chain reaches the true underlying graph in around $500$ iterations. We have also tried implementing Algorithm 1 with different initial graphs; all implementations resulted in the same posterior mode at the true underlying graph. 

We compare the performance of our approach with three other methods: the Gaussian graphical model of \citeasnoun{Jones2005} based on Metropolis-Hastings (GGM-MH), the graphical LASSO (GLASSO) of \citeasnoun{Friedman2008}, and the matrix-normal graphical model (MNGM) of \citeasnoun{Wang2009}. As both GGM-MH and GLASSO assume that each node is associated with one variable, we reduce the dimension of the functional data by retaining only the first principal component score. The MNGM method assumes matrix data, so we take the first five principal component scores and stack them up to form a $6\times 5$ matrix for each sample. In the MNGM method,  graph estimates across the rows and columns are obtained simultaneously, and only that across the rows is of interest to us.

The simulation results are demonstrated in the top panel of Table \ref{tab:simu1_2}. 
Summary statistics, such as running-time, mis-estimation rate, sensitivity and specificity are calculated for each method. The running-time was obtained using a laptop with Intel(R) Core(TM) i5 CPU, M430 with 2.27 GHZ processor and 4GB RAM.  The comparison of running-time shows that the GLASSO method is the fastest. This is because GLASSO does not require posterior sampling. However, GLASSO relies on a penalized optimization approach which requires determination of the tuning parameter. In this simulation, we have selected the tuning parameter that results in the lowest mis-estimation rate with respect to the underlying true graph. When the true graph is unknown, the tuning procedure can be time-consuming. The MNGM is much slower to implement, perhaps due to the numerical approximation of the marginal density in the MCMC algorithm.

In Table \ref{tab:simu1_2}, the mis-estimation rate is defined as the proportion of mis-estimated edges, obtained by averaging across all posterior samples. The sensitivity is the proportion of missed edges among the true edges, and the specificity is the proportion of over-estimated edges among the true non-edge pairs. The top panel of Table \ref{tab:simu1_2} shows that the proposed functional data graphical model provides the smallest mis-estimation rate as well as the highest sensitivity and specificity. We also observe that, although relying on excessive dimension reduction, the Gaussian graphical model and the GLASSO still provide reasonably good estimates. This suggests that for problems involving more nodes ($>$50), we can use these methods to obtain an initial estimate before applying our approach. 

\subsection{Simulation 2: Graph Estimation for Noisy Functional Data.}

We add white noise to the functional data generated in Simulation 1 to demonstrate the performance of posterior inference for noisy data. The variances of the additive white noise $\{\epsilon_{ij}(t)\}$ are generated from a gamma distribution with mean $2.5$ and variance $0.25$, resulting in a signal-to-noise ratio around $9$, where the signal-to-noise ratio is defined by $f_{ij}(t)/\mbox{var}\{\v_{ij}(t)\}$ and is averaged across the grid points and the samples. We apply model (\ref{postexpre}) and generate posterior samples using Algorithm 2. The eigenbasis and the variance of the noise are estimated simultaneously using the PACE algorithm. The principal component scores $\md_i$ are estimated by projecting the raw data on the estimated eigenbasis. The parameter $\boldsymbol{\Lambda}$ is determined using the estimated variance of the white noise, and the other model parameters are set to be the same as in Simulation 1. The posterior inference results are compared with the other three methods in the bottom panel of Table ~\ref{tab:simu1_2}. Similar patterns are observed as in Simulation 1. In particular, the proposed functional data graphical model shows a clear advantage in accurately estimating the graph. Estimates of the functions $\{ f_{ij} \}$ and their time-domain correlations are provided in the supplementary material.

\subsection{Simulation 3: Graph Estimation When p is Greater than n}

To further investigate the performance of the proposed approach when the number of nodes $p$ is greater than the sample size $n$, we design another simulation study with $p = 60$ and $n = 55$. The true graph contains $60$ nodes, among which $42$ are singletons and $18$ are connected with edges. The total number of edges in the true graph is $57$. Smooth functional data are simulated following the procedure described in Section \ref{sec:simu1}. With the simulated data, we apply the PACE algorithm to estimate $\{\mc_i\}$ and determine the truncation parameters using the FVE criterion with a 95\% threshod. We then apply Algorithm 1 and set prior parameters $\delta$ and ${\bf U}$ following Simulation 1. Posterior samples of the graph are obtained for $30,000$ MCMC iterations after removing $10,000$ burn-in samples.

\begin{figure}
\centering
\includegraphics[width=5 in,angle=0]{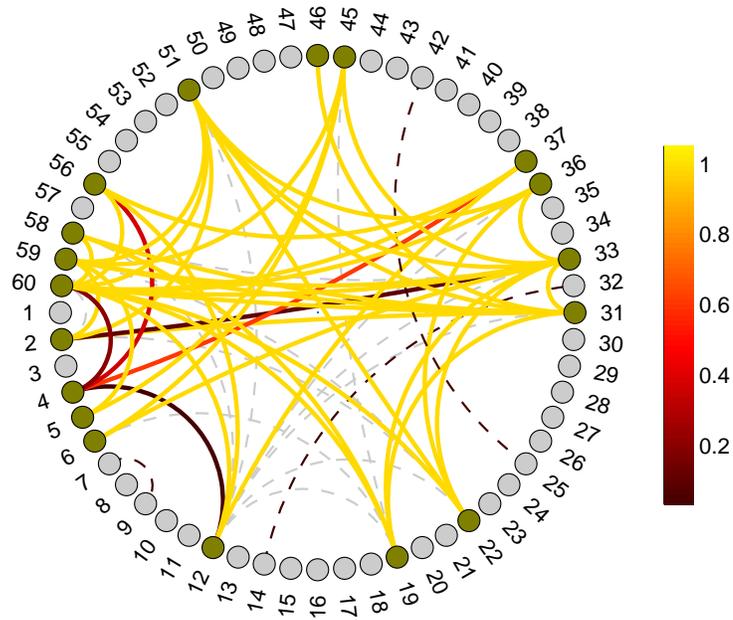}
\caption[Plot of Simulation 3]{Plot of Simulation 3: the estimated graph based on the marginal inclusion probability for each edge.}
\label{fig:largep}
\end{figure}

The posterior inference results are summarized in a circular graph plot in Figure \ref{fig:largep}, where we show an estimated graph by thresholding the marginal inclusion probability for each edge---the proportion that each edge is included in the posterior samples---to be greater than $0.03$. In Figure \ref{fig:largep}, the colors indicate the levels of the marginal inclusion probabilities, the colored dashed lines indicate edges that are mistakenly estimated, and the gray dashed lines indicate edges that are missed. This gives $46$ estimated edges, among which $43$ are correctly estimated, and $3$ are mistakenly estimated. Additionally, $14$ edges in the true graph are missed. We have also calculated the summary statistics similarly as in previous simulations, resulting in mean mis-estimation rate $0.01$, sensitivity $0.69$, and specificity $0.99$. Extra simulation runs show that the sensitivity level is improved when we increase the sample size $n$.

\section{Analysis of EEG data in an alcoholism study}
\label{sec5:realdata}

We apply the proposed method to EEG data from an alcoholism study. Data were obtained from 64 electrodes placed on subjects' scalps that captured EEG signals at 256 Hz during a one-second period.  The measurements were taken from 122 subjects, including 77 subjects who were in the alcoholism group and 45 in the control group. Each subject completed 120 trials. During each trial, the subject was exposed to either a single stimulus (a single picture) or two stimuli (a pair of pictures) shown on a computer monitor. We band-pass filtered the EEG signals to extract the $\alpha$ frequency band in the range of 8--12.5 Hz. The filtering was performed by applying the {\texttt{eegfilt}} function in the EEGLAB toolbox of Matlab. The $\alpha$-band signal is known to be associated with inhibitory control \cite{Knyazev2007}. 
Research has shown that, relative to control subjects, alcoholic subjects demonstrate unstable or poor rhythm and lower signal power in the $\alpha$-band signal \cite{Porjesz2005,Finn1999}, indicating decreased inhibitory control \cite{Sher2005}. Moreover, regional asymmetric patterns have been found in alcoholics---alcoholics exhibit lower left $\alpha$-band activities in anterior regions relative to right \cite{Hayden2006}. In this study, we aim to estimate the conditional independence relationships of $\alpha$-band signals from different locations of the scalp, and expect to find evidence that reflects differences in brain connectivity and asymmetric pattern between the two groups.

Since multiple trials were measured over time for each subject, the EEG measurements may not be treated as independent due to the time dependence of the trials. Furthermore, since the measurements were taken under different stimuli, the signals could be influenced by different stimulus effects. To remove the potential dependence between the measurements and the influence of different stimulus types, for each subject, we averaged the band-filtered EEG signals across all trials under the single stimulus, resulting in one Event-related potential (ERP) curve per electrode per subject. ERP is a type of electrophysiological signal generated by averaging EEG segments recorded under repeated applications of a stimulus, with the averaging serving to reduce biological noise levels and enhance the stimulus evoked neurological signal \cite{Brandeis1986,Bressler2002}. Based on the preprocessed ERP curves, we further removed subjects with missing nodes, and balanced the sample size across the two groups, producing multivariate functional data with $n=44$ and $p=64$ for both the alcoholic and the control group. We applied model~(\ref{eq:mod_1}) using coefficients of the eigenbasis expansion. The number of eigenbasis $\{M_j\}$ was determined through retaining $90\%$ of the total variation; this resulted in 4--7 coefficients per $f_j$. We collected $30,000$ posterior samples using Algorithm 1, in which the first $10,000$ were treated as the burn-in period. The model was fitted for both the alcoholic and the control group, and convergence of the MCMC was justified by running multiple chains starting with various initial values. 

\begin{figure}
\centering
\includegraphics[width=5.1 in, angle=0]{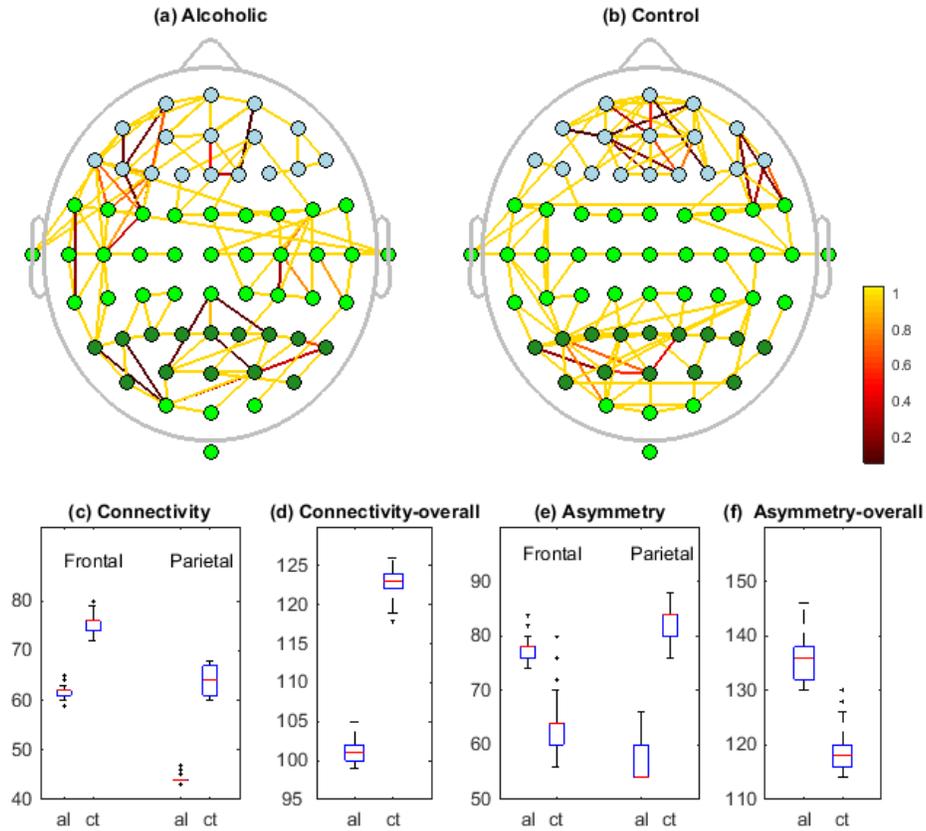}
\caption[Summary of posterior inference]{Summary of posterior inference: the marginal inclusion probabilities for edges in the alcoholic group (a) and the control group (b); the boxplots of connectivity measures: the number of edges connecting with nodes in the frontal and parietal regions (c), and the overall total number of edges (d); the boxplots of asymmetry measures: the number of asymmetric edges for nodes in the frontal and the parietal regions (e), and the overall total number of asymmetric edges (f). In (a) and (b), the edge color indicates the magnitude of the posterior inclusion probability. In (c)--(f), the alcoholic group is abbreviated as ``al", and the control group is abbreviated as ``ct".}
\label{fig:real1}
\end{figure}

The posterior results are summarized in Figure~\ref{fig:real1}.  The plots in (a) and (b) show the marginal inclusion probabilities for edges in the alcoholic and the control group respectively, where the edge color indicates the proportion that each edge is included in the posterior samples. To distinguish different regions, we used light blue to highlight nodes in the frontal region, used dark green to highlight nodes in the parietal region, and used green to indicate nodes in the central and occipital regions. Comparing (a) with (b), we see that the alcoholic group contains more edges connecting the left frontal-central, right central, and right parietal regions than the control group. The control group, on the other hand, contains more edges connecting the middle and right frontal regions, as well as the left parietal region than the alcoholic group.

To further compare with established results, we calculated two 
summary statistics for connectivity: the number of edges connected with nodes in a specific region, and the overall total number of edges. We also calculated two additional summary statistics for asymmetry: the number of asymmetric edges for all nodes in a specific region, and the overall total number of asymmetric edges. We summarized these summary statistics across the two groups using boxplots in Figure~\ref{fig:real1} (c)--(f), and calculated the posterior probability that the alcoholic group is greater than, equal to, or less than the control group for each statistic. Results show that, with probability $\approx 1$, the alcoholic group has fewer edges than the control group in the frontal and the parietal region, and has fewer overall total number of edges; with probability $0.95$, the alcoholic group has more asymmetric edges than the control group in the frontal region; and with probability $\approx 1$, the alcoholic group has higher overall total number of asymmetric edges than the control group. These results indicate that the alcoholic group exhibits decreased regional and overall connectivity, increased asymmetry in the frontal region, and increased overall asymmetry. These observations are consistent with the findings of \citeasnoun{Hayden2006}, who studied the asymmetric patterns at two frontal electrodes (F3, F4) and  two parietal electrodes (P3, P4) using the analysis of variance method based on the resting-state $\alpha$-band power. In comparison, our analysis provides connectivity and asymmetric pattern of all $64$ electrodes simultaneously whereas \citeasnoun{Hayden2006} only focuses on the four representative electrodes.

\section{Discussion}
\label{sec6:discus}

We have constructed a theoretical framework for graphical models of multivariate functional data and proposed a HIWP prior for the special case of Gaussian process graphical models. For practical implementation, we have suggested a posterior inference approach based on a regularization condition, which enables posterior sampling through MCMC algorithms. 

One concern is whether it is possible to perform exact posterior inference without the regularity condition on approximation, i.e., inferring the graph directly from the joint posterior $p(G|\{\mc_i\})\propto p(\{\mc_i\}| G)p(G)$
based on model (\ref{eq:mod_1}), where $p(\{\mc_i\}| G)$ is the marginal likelihood (with the covariance kernel $\mathcal{Q}_{\mathcal{C}}$ integrated out) and $p(G)$ is the prior distribution for $G$. Although the above joint posterior is theoretically well-defined according to Theorem 2, exact posterior sampling is difficult due to the fact that the density function for the marginal likelihood can only be calculated on a finite dimensional projection of $\{\mc_i\}$. 

In posterior inference, the influence of the approximation error on the posterior distribution can be quantified empirically.  Assuming that the functional data are pre-smoothed, the approximation error can be quantified by calculating the difference of the $\ell^2$ norms between the full sequence and the truncated sequence. The influence on the posterior distribution can be quantified by measuring the sensitivity of the posterior distribution to the change of truncation \cite{saltelli2000}. For example, based on model (\ref{eq:mod_1}) one may calculate the Kullback-Leibler divergence  
for two different truncation parameters $M$ and $M'$. An alternative method for pre-determining the truncation parameter is to choose a prior for $M$ in a Bayesian hierarchical model, in which case hybrid MCMC algorithms are needed for fitting both models (\ref{eq:mod_1}) and (\ref{postexpre}). The posterior sampling in these models would become more complicated because the dimension of the truncated sequences and the size of the covariance matrix ${\bf Q}_{\mathcal{C}}$ would change whenever $M$ is updated.

We have focused on decomposable graphs. In case of non-decomposable graphs, the proposed $\mbox{HIWP}$ prior may still apply if we replace the inverse-Wishart process prior for each clique with that for a prime component of the graph. For a non-complete prime component $P$, the inverse-Wishart processes prior for $\mathcal{\bf Q}_P$ is subject to extra constraint induced by missing edges.

We have applied the proposed method to graphs of small to moderate size, with number of nodes as large as $60$.
To deal with larger scale problems (e.g, multivariate functional data with hundreds or thousands of functional components),  more efficient large-scale computational techniques such as the fast Cholesky factorization \cite{li2012} can be readily combined with our MCMC algorithms. Furthermore, non-MCMC algorithms may be more computationally efficient in case of large graphs. For example, based on the posterior distribution of $G$ in (\ref{eq:mod2}), a fast search algorithm may be developed to search for the maximum a posteriori (MAP) solution following ideas similar to \citeasnoun{DaumeIII2007} and \citeasnoun{Jalali2011}.

\section*{Appendix: Proofs}

\subsection*{A. Definitions}
Definitions used in the lemmas, theorems and their proofs are listed as follows: {(I) \it Projection map.} Let $\mathbb{R}$ be the real line and $T$ be an index set. Consider the Cartesian product space $\mathbb{R}^{T\times T}=\prod_{(\alpha,\beta)\in T\times T}\mathbb{R}^{(\alpha,\beta)}$. For a fixed point $(\alpha,\beta) \in T\times T$, we define the projection map $\pi_{(\alpha,\beta)}: \mathbb{R}^{T\times T} \rightarrow \mathbb{R}^{(\alpha,\beta)}$ as $\pi_{(\alpha,\beta)}\left( \{ x_{(l,m)}:(l,m)\in T\times T\}\right)=x_{(\alpha,\beta)}$. For a subset $B\subset T\times T$,  we define the partial projection $\pi_{B}: \mathbb{R}^{T\times T} \rightarrow \mathbb{R}^{B}$ as $\pi_{B}\left( \{ x_{(l,m)}:(l,m)\in T\times T\}\right)=\{x_{(s,t)}: (s,t)\in B\}$.
More generally, for subsets $B_1, B_2$, such that $B_2 \subset B_1 \subset T\times T$, we define the partial sub-projections $\pi_{B_2 \leftarrow B_1}: \mathbb{R}^{B_1} \rightarrow \mathbb{R}^{B_2}$, by $\pi_{B_2\leftarrow B_1}(\,\{x_{(l,m)}: (l,m) \in B_1\}\,)=\{x_{(s,t)}: (s,t)\in B_2\}$. 
{(II) \it The pullback of a $\sigma$-algebra.} Let $\mathcal{B}_{(\alpha,\beta)}$ be a $\sigma$-algebra on $\mathbb{R}^{(\alpha,\beta)}$. We can create a $\sigma$-algebra on $\mathbb{R}^{T\times T}$ by pulling back the $\mathcal{B}_{(\alpha,\beta)}$ using the inverse of the projection map and define $\pi_{(\alpha,\beta)}^* (\mathcal{B}_{(\alpha,\beta)})=\{\pi_{(\alpha,\beta)}^{-1}(A): A \in \mathcal{B}_{(\alpha,\beta)}\}$. One can verify that $\pi_{(\alpha,\beta)}^* (\mathcal{B}_{(\alpha,\beta)})$ is a $\sigma$-algebra. 
{(III) \it Product $\sigma$-algebra.} We define the product $\sigma$-algebra as $\mathcal{B}(\mathbb{R}^{T\times T})=\prod_{(\alpha,\beta) \in T \times T} \mathcal{B}_{(\alpha,\beta)}$, where
$\prod_{(\alpha,\beta) \in T \times T} \mathcal{B}_{(\alpha,\beta)} = {\sigma}\left(\bigcup_{(\alpha,\beta)\in T\times T} \pi_{(\alpha,\beta)}^*(\mathcal{B}_{(\alpha,\beta)})\right).$ 
{(IV) \it Pushforward measure.} Given a measure $\mu_{T\times T}$ on the product $\sigma$-algebra, and a subset  $B$ of $T\times T$, we define the pushforward measure $\mu_{B}=(\pi_{B})_* \mu_{T\times T}$ on $\mathbb{R}^{B}$ as   
$\mu_{B}(A)=\mu_{T\times T}\{\pi_{B}^{-1}(A)\}$
for all $A\in \mathcal{B}_{B}$, where $\mathcal{B}_{B}=\prod_{(\alpha,\beta)\in B}\mathcal{B}_{(\alpha,\beta)}$.
{(V) \it Compatibility.} Given subsets $B_1,B_2$ of $T\times T$ such that $B_2 \subset B_1 \subset T\times T$, the pushforward measures $\mu_{B_1}$ and $\mu_{B_2}$ are said to obey compatibility relation if 
$(\pi_{B_2 \leftarrow B_1})_*\mu_{B_1}=\mu_{B_2}$.

\subsection*{B. Proof of Lemma 1}

This proof involves some measure-theoretic arguments. The essential idea is to use disintegration theory \cite{Chang1997} to first construct the conditional probability measure $P_1\left\{\cdot \mid \pi_{A\cap B}(\mf_A)\right\}$ on $\mathcal{B}(L^2(T_A))$, extend this to $P\{\:\cdot\:\mid \pi_B(\mf)\}$ on $\mathcal{B}(L^2(T_{A\cup B}))$, and finally construct the joint measure $P$ which satisfies conditions (i)--(iii).

Denote $T_A=\bigsqcup_{j\in A} T_j$. Since $P_1$ is a finite Radon measure and the projection $\pi_{A\cap B}: L^2(T_A)\rightarrow L^2(T_{A\cap B})$ is measurable, we invoke the disintegration theorem to obtain measures $P_1\left\{\cdot \mid \pi_{A\cap B}(\mf_A)\right\}$ on $\mathcal{B}(L^2(T_A))$ satisfying: \,(a.1)\, $\displaystyle P_1(\mathcal{X}\mid \mf_{A\cap B})=P_1\left\{\mathcal{X}\cap[L^2(T_{A\setminus B})\times\{\pi_{A\cap B}(\mf_A)\}]\mid \pi_{A\cap B}(\mf_A)\right\}$ for all $\mathcal{X}\in\mathcal{B}(L^2(T_A))$, 
\, (b.1)\, the map $\displaystyle \mf_{A\cap B}\mapsto (P_1)_{\mf_{A\cap B}} H \:\colon\hspace{-1px}=\int H(\mf_A) dP_1(\mf_A\mid \mf_{A\cap B})$ is measurable for all nonnegative measurable $H: L^2(T_A)\rightarrow \mathbb{R}$, 
and \,(c.1)\,  $\displaystyle P_1 H =((\pi_{A\cap B})_\ast P_1)(P_1)_{\mf_{A\cap B}} H$ for all nonnegative measurable $H: L^2(T_A)\rightarrow \mathbb{R}$, where $(\pi_{A\cap B})_\ast P_1$ is the push-forward measure of $P_1$.  

Now, we define the measure $P\{\:\cdot\:\mid \pi_B(\mf)\}$ by setting
$P\{\mathcal{A}\mid\pi_B(\mf)\}=P_1\{\pi_A(\mathcal{A}\cap [L^2(T_{A\setminus B})\times\{\pi_B(\mf)\}])\mid \pi_{A\cap B}(\mf)\}.$
Note that this is well defined for all measurable $\mathcal{A}\in\mathcal{B}(L^2(T_{A\cup B}))$ since the sections $\pi_A(\mathcal{A}\cap [L^2(T_{A\setminus B})\times\{\pi_B(\mf)\}])$ are always measurable, and also that \,(a)\,
$\displaystyle P\{\mathcal{A}\mid \pi_B(\mf)\}=P\{\mathcal{A}\cap[L^2(T_{A\setminus B})\times\{\pi_{B}(\mf)\}]\mid \pi_{B}(\mf)\}$
holds by construction. Now, let $\mathcal{M}$ denote the set of measurable functions from $L^2(T_{A\cup B})$ to $\mathbb{R}$ satisfying \,(b)\,
$\displaystyle \mf_B\longmapsto P_{\mf_B} H$ is a measurable function on $L^2(T_B)$. 
We shall argue that $\mathcal{M}$ is a monotone class. First, suppose $H_n$ is a sequence of positive measurable functions in $\mathcal{M}$ increasing pointwise to a bounded measurable function $H$. For each fixed $\mf_B$ in $L^2(T_B)$, we then have that $H_n$ is a sequence of positive measurable functions increasing pointwise to $H$, and hence the monotone convergence theorem implies
$P_{\mf_B} H_n\longrightarrow P_{\mf_B} H$ 
in an increasing manner. Since this holds for each $\mf_B$, we conclude that $P_{\mf_B} H$ is the point-wise increasing limit of measurable functions on $L^2(T_B)$, and hence it is measurable. Moreover, it is simple to see that
$P_{\mf_B} {\bf 1}_{\mathcal{X}\times \mathcal{Y}}=P_1(\mathcal{X}\mid \mf_{A\cap B}){\bf 1}_{\mathcal{Y}}(\mf_{B\setminus A})$
is a measurable function on $L^2(T_B)$ for all $\mathcal{X}\in\mathcal{B}(L^2(T_A))$ and $\mathcal{Y}\in\mathcal{B}(L^2(T_{B\setminus A}))$, and hence ${\bf 1}_{\mathcal{X}\times\mathcal{Y}}\in\mathcal{M}$. By the Monotone Class Theorem, we then have that all bounded measurable functions on $L^2(T_{A\cup B})$ satisfy (b), and hence it will hold for all positive measurable functions on $L^2(T_{A\cup B})$. 
Since (b) is satisfied for all positive measurable functions, we may define the measure
$P H=P_2 P_{\mf_B}H.$
By construction, we have that
$
 P{\bf 1}_{L^2(T_{A\setminus B})\times\mathcal{Y}}=P_2 P_1(L^2(T_{A\setminus B})\times \{\mf_{A\cap B}\}\mid \mf_{A\cap B}){\bf 1}_{\mathcal{Y}}(\mf_{B})=P_2(\mathcal{Y})
$
and
$ P{\bf 1}_{\mathcal{X}\times L^2(T_{B\setminus A})}= P_2 P_1(\mathcal{X}\mid\mf_{A\cap B})
=((\pi_{A\cap B})_\ast P_2)P_1(\mathcal{X}\mid \mf_{A\cap B})
=((\pi_{A\cap B})_\ast P_1)P_1(\mathcal{X}\mid \mf_{A\cap B})=P_1(\mathcal{X}).
$
Thus, we also have that
$P H= P_2 P_{\mf_B} H=((\pi_B)_\ast P) P_{\pi_B(\mf)} H$
for all measurable $H$, and this is the final property establishing that $P(\:\cdot\:\mid \mf_B)$ is a disintegration of $P$ with respect to the map $\pi_B$. By the disintegration theorem, this disintegration is a version of the regular conditional probability of $\mf_A$ given $\mf_B$. Since this version only depends upon $\mf_{A\cap B}$, we conclude that (iii) holds. Finally, we note that any other measure satisfying these properties must agree with the measure we have constructed on $\pi$-system, and therefore the uniqueness of $P$ immediately follows. \qedwhite

\subsection*{C. Proof of Proposition 1}

{\it Proof.} The Properties 1 - 4 in \citeasnoun{Dawid1993} are treated as axioms; they are universal properties thus also hold when $X, Y, Z$ are random processes. Since the graph $G$ is undirected and decomposable, the results on graphical theory in Appendix A of \citeasnoun{Dawid1993} continue to hold. Properties 1 - 4 and results in Appendix A imply that results in B1- B7 of \citeasnoun{Dawid1993} continue to hold when P is a Markov distribution constructed in Lemma 1. Theorem 2.6 and Corollary 2.7 of \citeasnoun{Dawid1993} are also implied. These results, combined with the definition of marginal distribution defined by pushforward measure and the definition of conditional probability measure based on disintegration theory, prove that Lemmas 3.1, 3.3, Theorems 3.9 - 3.10 as well as Propositions 3.11, 3.13, 3.15, 3.16, 3.18 from \citeasnoun{Dawid1993} hold. \qedwhite

\subsection*{D. Lemma 2 and proof}

{\it Lemma 2.} Let $\mathbb{N}$ be the set of positive integers and $I$ an arbitrary finite subset of it. Suppose that $\delta>4$ is a positive integer and that $u:\mathbb{N}\times \mathbb{N}\rightarrow \mathbb{R}$ is a symmetric positive semidefinite and trace class kernel so that the matrix $\bUI$ formed by $\{u(i,j), i,j \in I\}$ is symmetric positive semidefinite. Then there exists a unique probability measure $\mu$ on $(\mathbb{R}^{\mathbb{N}\times\mathbb{N}},\mathcal{B}(\mathbb{R}^{\mathbb{N}\times\mathbb{N}}))$ satisfying\\
\indent i.\,\, $(\pi_{I\times I})_\ast\mu=\mu_{I\times I}$,
where $\mu_{ I\times I}$ is the law of ${\mbox{\small IW}}(\delta, \bUI)$ defined in \citeasnoun{Dawid1981};\\
\indent ii.\,\, if $B=\{(\alpha_i,\beta_i)\}_{i=1}^n\subset \mathbb{N}\times \mathbb{N}$ and $\mathrm{g}=\{\alpha_i\}_{i=1}^n\cup\{\beta_i\}_{i=1}^n$, then $(\pi_B)_\ast\mu=\mu_B$, where $\mu_B=(\pi_{B\leftarrow \mathrm{g}\times \mathrm{g}})_\ast\mu_{\mathrm{g}\times \mathrm{g}}$.\\
Setting $\mu=\iwp(\delta, {\bf U})$ so that $({\bf U})_{ij}=u(i,j)$, we further have that if ${\bf Q}\sim \iwp(\delta, {\bf U})$ and $\delta>4$, the countably infinite array ${\bf Q}$ is a positive semidefinite trace class operator on $\ell^2(\mathbb{N})$ almost surely.

{\it Proof. } Let $\bUI$ be a matrix with the law $\mu_{I\times I}$. We will prove following \citeasnoun[Theorem 2.4.3]{Tao2011} as follows:\quad(1)\, we verify the compatibility of $\mu_B$ for all finite $B\subset \mathbb{N} \times \mathbb{N}$. There are two successive cases we shall consider. Case 1: Suppose $I_2\subset I_1$ are two finite subsets of $\mathbb{N}$, then ${\bf Q}_{I_2\times I_2}$ is the sub-matrix of ${\bf Q}_{I_1\times I_1}$ obtained by deleting the rows and columns with indices in $I_1\setminus I_2$. 
If ${\bf Q}_{I_1\times I_1}$ has law $\mu_{I_1\times I_1}={\mbox{ \small IW}}(\delta,{\bf U}_{I_1\times I_1})$, then ${\bf Q}_{I_2\times I_2}$ has law $\iw(\delta, {\bf U}_{I_2\times I_2})$ due to the consistency property of the inverse-Wishart distribution \cite[Lemma 7.4]{Dawid1993}.  Consequently, 
$(\pi_{I_2\times I_2\leftarrow I_1\times I_1})_\ast\mu_{I_1\times I_1} =\mu_{I_2\times I_2}.$ 
Case 2: Let $B_1=\{(\alpha_i,\beta_i)\}_{i=1}^n\subset \mathbb{N}\times \mathbb{N}$ and suppose $B_2=\{(\widetilde{\alpha}_i,\widetilde{\beta}_i)\}_{i=1}^m\subset B_1$. Set $\mathrm{g}_1=\{\alpha_i\}_{i=1}^n\cup\{\beta_i\}_{i=1}^n$ and $\mathrm{g}_2=\{\widetilde{\alpha}_i\}_{i=1}^m\cup\{\widetilde{\beta}_i\}_{i=1}^m$ so that $\mathrm{g}_2\times \mathrm{g}_2\subset \mathrm{g}_1\times \mathrm{g}_1$. It is clear that
$\pi_{B_2\leftarrow B_1}\circ\pi_{B_1\leftarrow \mathrm{g}_1\times \mathrm{g}_1}=\pi_{B_2\leftarrow \mathrm{g}_1\times \mathrm{g}_1}=\pi_{B_2\leftarrow \mathrm{g}_2\times \mathrm{g}_2}\circ\pi_{\mathrm{g}_2\times \mathrm{g}_2\leftarrow \mathrm{g}_1\times \mathrm{g}_1}.$
Thus, 
\begin{align*}
(\pi_{B_2\leftarrow B_1})_\ast\mu_{B_1}&=(\pi_{B_2\leftarrow B_1})_\ast(\pi_{B_1\leftarrow \mathrm{g}_1\times \mathrm{g}_1})_\ast\mu_{\mathrm{g}_1\times \mathrm{g}_1}
=(\pi_{B_2\leftarrow B_1}\circ\pi_{B_1\leftarrow \mathrm{g}_1\times \mathrm{g}_1})_\ast\mu_{\mathrm{g}_1\times \mathrm{g}_1} \nonumber\\
&=(\pi_{B_2\leftarrow \mathrm{g}_2\times \mathrm{g}_2}\circ\pi_{\mathrm{g}_2\times \mathrm{g}_2\leftarrow \mathrm{g}_1\times \mathrm{g}_1})_\ast\mu_{\mathrm{g}_1\times \mathrm{g}_1}
=(\pi_{B_2\leftarrow \mathrm{g}_2\times \mathrm{g}_2})_\ast(\pi_{\mathrm{g}_2\times \mathrm{g}_2\leftarrow \mathrm{g}_1\times \mathrm{g}_1})_\ast\mu_{\mathrm{g}_1\times \mathrm{g}_1}\nonumber\\
&=(\pi_{B_2\leftarrow \mathrm{g}_2\times \mathrm{g}_2})_\ast\mu_{\mathrm{g}_2\times \mathrm{g}_2}=\mu_{B_2},
\end{align*}
where the second to last equality holds because of our demonstration in Case 1.
\,(2) \,Second, we claim that the finite dimensional measure $\mu_{I\times I}=\iw(\delta,{\bf U}_{I \times I})$ is an inner regular probability measure on the product $\sigma$-algebra $\mathcal{B}_{I \times I}$. We will show that $\mu_{I \times I}$ is a finite Borel measure on a Polish space, which then implies that $\mu_{I\times I}$ is regular, hence inner regular by \citeasnoun[Lemma 26.2]{bauer2001}. This is done through (a)--(c) as follows:
(a) For finite $I$, ${\bf Q}_{I\times I}$ takes values in the space of symmetric and positive semidefinite matrices, denoted by ${\Psi}_{\vert I\vert}$ where $\vert N\vert$ denotes the number of elements in $I$.  Since the subset of symmetric matrices is closed in $\mathbb{R}^{I\times I}$, it is Polish. Furthermore, the space of symmetric positive semidefinite matrices is an open convex cone in the space of symmetric matrices, hence it is Polish as well. Therefore the space $\Psi_{\vert I\vert}$ is Polish.
(b) Since $\mu_{I \times I}$, the law of  ${\bf Q}_{I\times I} \sim \iw(\delta, U_{I \times I})$, has an almost everywhere continuous density function, $\mu_{I \times I}$ is a measure defined by Lebesgue integration against an almost everywhere continuous function. Therefore $\mu_{I \times I}$ is Borel on $\Psi_{\vert I\vert}$. As $\Psi_{\vert I\vert}\subset \mathbb{R}^{I\times I}$, we may extend the measure $\mu_{I \times I}$ from $\Psi_{\vert I\vert}$ to $\mathbb{R}^{I\times I}$ via the Carath\'eodory theorem \cite[Theorem 1.7.3]{Tao2011}. In particular,  define $\widetilde{\mu}_{I \times I}(A)={\mu}_{I \times I}(A\cap \Psi_{\vert I\vert })$ for $A\in \mathcal{B}(\mathbb{R}^{I\times I})$. With extension, $\mu_{I \times I}$ is Borel on $\mathbb{R}^{I\times I}$, and the $\sigma$-algebra associated is $\mathcal{B}(\mathbb{R}^{I\times I})=\mathcal{B}_{I \times I}=\prod_{(\alpha,\beta)\in I\times I}\mathcal{B}_{(\alpha,\beta)}$.       
\,(c)  The measure $\mu_{I\times I}$ is certainly finite since it is a probability measure. 

The compatibility and regularity conditions in (1) and (2) ensure that the Kolmogorov extension theorem holds. Therefore there exists a unique probability measure $\mu$ on the product $\sigma$-algebra $\mathcal{B}(\mathbb{R}^{\mathbb{N}\times \mathbb{N}})$ that satisfies (i) and (ii).

We now prove that if ${\bf Q} \sim \iwp(\delta, {\bf U})$, then the countably infinite array ${\bf Q}$  is a well-defined positive semidefinite trace class operator on $\ell^2(\mathbb{N})$ almost surely. First, we note that the spectral theorem ensures the existence of an orthonormal basis of $\ell^2(\mathbb{N})$ that diagonalizes $U$.  Thus, without loss of generality, we may assume that ${\bf Q}$ is drawn from $\iwp(\delta,{\bf U})$ where ${\bf U}$ is a diagonal positive semidefinite trace class operator on $\ell^2(\mathbb{N})$.  

First, we show each row of ${\bf Q}{\bf x}$ is finite almost surely hence is well-defined for all ${\bf x}\in\ell^2(\mathbb{N})$. It is sufficient to show that $E[\,|({\bf Qx})_i|\,]<\infty$. We note that for arbitrary $i\neq j$,
$\left(\begin{array}{rr}
			q_{ii}&q_{ij}\\
			q_{ij}&q_{jj}
			\end{array}\right)\sim \iw\left(\delta,\left(\begin{array}{rr}
														u_{ii}&0\\
														0&u_{jj}
														\end{array}\right)\right)$
and hence using the moments of finite dimensional inverse-Wishart, 
$E (q_{ii}^2)= u_{ii}^2(\delta-2)^{-1}(\delta-4)^{-1}, E (q_{ij}^2)=u_{ii}u_{jj}(\delta-1)^{-1}(\delta-2)^{-1}(\delta-4)^{-1}$,  for $\delta>4$.
By Tonelli's theorem, we have that
$E \sum_j q_{ij}^2=\sum_j E q_{ij}^2\leq C\sum_j u_{ii}u_{jj}=Cu_{ii}\sum_j u_{jj}$,
where $C$ is the maximum of the above constants.
Thus $E[\,|({\bf Qx})_i|\,]\leq \Vert {\bf x} \Vert \sqrt{E\sum_j q_{ij}^2}<\infty$. 
Because there are only countably many rows, we have that ${\bf Qx}$ is finite almost surely for all rows simultaneously. Consequently, we have that ${\bf Qx}$ is well-defined for all ${\bf x}\in\ell^2(\mathbb{N})$. Now we show that ${\bf Qx} \in \ell^2(\mathbb{N})$ almost surely. By similar considerations, let ${\bf q}_i = ({\bf Qx})_i$, then 
$E(\sum_i \Vert {\bf q}_i\Vert^2)\leq C\left(\sum_i u_{ii}\right)^2<\infty$ and $\Vert {\bf Qx}\Vert^2\leq C\Vert {\bf x}\Vert^2\sum_i \Vert q_i\Vert^2$; this implies that $\Vert {\bf Qx} \Vert <\infty$ almost surely hence ${\bf Qx} \in \ell^2(\mathbb{N})$ almost surely, and it also implies that the operator norm $\Vert {\bf Q}\Vert_{op}$ is finite almost surely.

By construction, we must have that ${\bf Q}$ is positive semidefinite almost surely since
$\langle {\bf Qx},{\bf x}\rangle=\lim_{n\rightarrow\infty}\langle {\bf Q}_n {\bf x},{\bf x}\rangle\geq 0,$
where ${\bf Q}_n$ is the restriction of ${\bf Q}$ to its $n$ by $n$ leading principal minor. Finally, ${\bf Q}$ is trace class almost surely since
$E[\,|\mbox{tr}({\bf Q})|\,]=\sum_i E(q_{ii}) =(\delta-2)^{-1}\sum_i u_{ii} <\infty$.
\qedwhite 

\subsection*{E. Proof of Theorem 1}

{\it Proof. } Based on Lemma 2, we can define a sequence of inverse-Wishart process prior for $\mathcal{Q}_C$, denoted by $\mathcal{Q}_C \sim \iwp (\delta, \mathcal{U}_C), C\in \mathcal{C}$. These sequences are pairwise consistent due to the consistency of inverse-Wishart processes and the fact that $\mathcal{U}_{\mathcal{C}}$ is a common collection of kernels. Therefore, we can construct a unique hyper Markov law for $\mathcal{Q}_{\mathcal{C}}$ following procedure (12) - (13) of \citeasnoun{Dawid1993}. And Theorem 3.9 of \citeasnoun{Dawid1993} guarantees that the constructed hyper Markov law is unique. \qedwhite

\subsection*{F. Proof of Proposition 2}

{\it Proof. } Note that an operator drawn from a hyper-inverse-Wishart process with the parameter $\mathcal{U}$ satisfies \mbox{rank}$(u_{ij}) <\infty$ for $i, j \in V$ will have finite-rank almost surely. This follows by noting that if $\mathcal{Q}\sim \hiwp(\delta,\mathcal{U})$ and $\mathcal{W}$ is a fixed unitary transformation on $\ell^2$, then
$\mathcal{W}^T \mathcal{Q}\mathcal{W}\sim \hiwp(\delta,\mathcal{W}^T \mathcal{U}\mathcal{W})$. Thus, choosing $\mathcal{W}$ so that the block representation
$\mathcal{W}^T \mathcal{U}\mathcal{W} = \begin{pmatrix}
U & 0 \\
0 & 0
\end{pmatrix}$
holds (here, $U$ is a finite matrix and $0$'s represent infinite arrays of zeros), we see that the block representation
$\mathcal{W}^T \mathcal{Q}\mathcal{W} = \begin{pmatrix}
Q & 0 \\
0 & 0
\end{pmatrix}$
holds almost surely, and that $Q\sim \iw(\delta,U)$. Consequently, we have reduced to the finite-dimensional setting where the result is well-known. \qedwhite

\subsection*{G. Proof of Theorem 2}

{\it Proof. } By the result of Proposition 1, the $\hiwp_G$ prior is a strong hyper Markov law. So by Corollary 5.5 of \citeasnoun{Dawid1993}, the posterior law of $\mathcal{Q}_{\mathcal{C}}$ is the unique hyper Markov law specified by the marginal posterior laws at each clique. In other words, we just need to find the posterior law for the model: $\mc_{i,C}\sim \dmgp(\mc_{0,C}, \mathcal{Q}_{C})$ with prior $\mathcal{Q}_C \sim \iwp (\delta, \mathcal{U}_{C})$ for each $\mathcal{Q}_{C}$, and use them to construct the posterior law of $\mathcal{Q}_{\mathcal{C}}$ following (12) - (13) of \citeasnoun{Dawid1993}. As in the last proof, choosing an appropriate transformation reduces this to the finite-dimensional case which is well-known. Finally, by Proposition 5.6 of \citeasnoun{Dawid1993}, the marginal distribution of $\{\mathrm{c}_i\}$ given $G, \mc_0, \delta,  \widetilde{\mathcal{U}}_{\mathcal{C}}$ is again Markov over $G$. \qedwhite

\section*{Supplementary Materials}

 The supplementary document contains more detailed derivations, discussions, and simulation results.

\setlength{\baselineskip}{15 pt}
\bibliographystyle{ECA_jasa}
\bibliography{graph_refer}

\begin{table}
\centering
\caption[Summary statistics and comparison with three other methods]{Summary statistics of Simulation 1 and 2. 
nFPC: the number of functional principal components used to approximate each curve; Time: the running time (in seconds) based on $5000$ MCMC iterations; nEdge: the  total number of edges of the graph averaged across all posterior samples; nUnique: the number of unique graphs visited after the burnin period; MisR: the mean mis-estimation rate with respect to the true graph; Sen: sensitivity; Spec: specificity; FDGM-S: the proposed functional data graphical model for smooth data, based on Algorithm 1; FDGM-N: the proposed functional data graphical model for noisy data, based on Algorithm 2; GGM-MH: Gaussian graphical model; GLASSO: graphical LASSO; MNGM: matrix-normal graphical model.
} \label{tab:simu1_2} \vspace{2mm} 
\begin{tabular}{*{9}{lcccccccc}} \hline \hline
Data & Method    & nFPC      & Time                      &  nEdge  & nUnique     & MisR & Sen & Spec   \\\hline
         & FDGM-S     & 3 - 5          &  38~~                                    &   7.66      &    ~3           &   0.02   &  0.96 & 1.0~ \\  
 Smooth  & GGM-MH     & 1              &  0.15                                &    9.55     &    63          &    0.10    &   1.0~   & 0.78\\               
  & GLASSO     & 1              &  -                                           &   -                &     -    &  0.13 & -  & -\\
         & MNGM       & 5              &  4067.73                                &   5.83       &     36          &  0.21 &  0.66 & 0.93\\\hline 
         & FDGM-N     & 3 - 5          &64                                 &  7.86  &     ~5               &  0.01 & 0.98~  & 1.0~\\
Noisy    & GGM-MH     & 1              &  0.39                                   &   9.62     & 59     &    0.11  &  1.0~ & 0.77\\
  & GLASSO     & 1              & -                                           & - &         -            & 0.13 & - & -\\
         & MNGM       & 5              &  4086.38                            &   6.33    &        18          &  0.26 & 0.65 & 0.85\\  \hline          
\end{tabular}
\end{table}

\newpage
\baselineskip=25 pt
{\Large
\begin{center}
\title{Supplementary Materials for ``Bayesian Graphical Models for Multivariate Functional Data"}

\author{\normalsize Hongxiao Zhu$^{1}$, Nate Strawn$^{2}$, and David B. Dunson$^{3}$\\ [.05em]
\small $^1$ Virginia Tech, Blacksburg, VA 24061\\ [.01em]
\small $^2$ Georgetown University, Washington, DC 20057\\ [.01em]
\small $^3$ Duke University, Durham, NC 27708\\ [.01em]}
\end{center}
\date{}
\maketitle
}
\section*{1. More details of Algorithm 1}  
\begin{itemize}[leftmargin=0 pc]\itemsep0em
\item[]{\it Step 0.}{ Choose an initial decomposable graph $G$ and the prior parameters $\mathrm{c}_0$, $\delta$, ${\bf U}$. }
\item[]{\it Step 1.} With probability $1-p$, propose $\widetilde{G}\mid G \sim p(\widetilde{G}\mid G)$ by randomly adding or deleting an edge (each with probability $0.5$) in the space of decomposable graphs, and accept the new $\widetilde{G}$ with probability
\[\alpha=\min\left \{1, \frac{p(\widetilde{G}\mid \{{\mc}_i^M\},\mc_0^M)\; p(G\mid \widetilde{G})}{p(G\mid \{{\mc}_i^M\},\mc_0^M)\; p(\widetilde{G}\mid G)}\right\}.\]
For the case of adding (i.e. $\widetilde{G}$ has one more edge than $G$), there are two cases. Case (1),  the two nodes (denoted as $k,l$) being connected belong to two different connected components. Here a connected component is defined as a cluster of nodes that are connected so that for any node in the cluster there is a route from one node to another. In this case, the likelihood ratio takes the form:
\begin{align*} 
\frac{ p(\{{\mc}_i^M\}\mid {\mc}_0^M, \widetilde{G})}{ p(\{{\mc}_i^M\}\mid {\bf c}_0^M, G)}=&\frac{|{\bf U}_{k,l}|^{(\delta+d_{k,l}-1)/2}}{|{\bf U}_{k,k}|^{(\delta+d_{k,k}-1)/2}|{\bf U}_{l,l}|^{(\delta+d_{l,l}-1)/2}}\\
&\times \frac{|\widetilde{\bf U}_{k,k}|^{(\widetilde{\delta}+d_{k,k}-1)/2}|\widetilde{\bf U}_{l,l}|^{(\widetilde{\delta}+d_{l,l}-1)/2}}{|\widetilde{\bf U}_{k,l}|^{(\widetilde{\delta}+d_{k,l}-1)/2}}\\
&\times\frac{\Gamma_{d_{k,l}}(\frac{\widetilde{\delta}+d_{k,l}-1}{2})}{\Gamma_{d_{k,l}}(\frac{\delta+d_{k,l}-1}{2})} \frac{\Gamma_{d_{k,k}}(\frac{\delta+d_{k,k}-1}{2})}{\Gamma_{d_{k,k}}(\frac{\widetilde{\delta}+d_{k,k}-1}{2})} 
 \frac{\Gamma_{d_{l,l}}(\frac{\delta+d_{l,l}-1}{2})}{\Gamma_{d_{l,l}}(\frac{\widetilde{\delta}+d_{l,l}-1}{2})},
\end{align*}
where ${\bf U}_{k,k}$, ${\bf U}_{l,l}$ and ${\bf U}_{k,l}$ are sub-matrices of ${\bf U}$ associated with corresponding functional components, and $\Gamma_d(a)=\pi^{d(d-1)/2}\prod_{i=0}^{d-1}\Gamma(a-i/2)$. Here $d_{k,k}$, $d_{l,l}$ and $d_{k,l}$ are the size of the corresponding sub-matrices.
Case (2), the two nodes $k,l$ being connected belong to the same connected components. The decomposability implies that after connecting, $k,l$ lie in the same clique, denoted as $C_q$. Denote $S_q=C_q\setminus \{k,l\}$, 
$C_{q_1}=C_q\setminus k$, $C_{q_2}=C_q\setminus l$ and $D=\{k,l\}$, we can write ${\bf U}_{C_q}$ in the form of 
\begin{eqnarray*}
 \begin{pmatrix}
  {\bf U}_{S_q}   & {\bf U}_{S_q,D}  \\
  {\bf U}_{D,S_q} & {\bf U}_D 
 \end{pmatrix}. 
\end{eqnarray*}
Then the likelihood ratio takes the form 
\begin{align*} 
\frac{ p(\{{\mc}_i^M\}\mid {\mc}_0^M, \widetilde{G})}{ p(\{{\mc}_i^M\}\mid {\mc}_0^M, G)}=&\frac{|\bU_{C_q}|^{(\delta+d_{C_q}-1)/2}|\bU_{S_q}|^{(\delta+d_{S_q}-1)/2}}{|\bU_{C_{q_2}}|^{(\delta+d_{C_{q_2}}-1)/2}|\bU_{C_{q_1}}|^{(\delta+d_{C_{q_1}}-1)/2}}\\
&\times
\frac{|\widetilde{\bU}_{C_{q_2}}|^{(\widetilde{\delta}+d_{C_{q_2}}-1)/2}|\widetilde{\bU}_{C_{q_1}}|^{(\widetilde{\delta}+d_{C_{q_1}}-1)/2}}{|\widetilde{\bU}_{C_q}|^{(\widetilde{\delta}+d_{C_q}-1)/2}|\widetilde{\bU}_{S_q}|^{(\widetilde{\delta}+d_{S_q}-1)/2}}\\
&\times \frac{\Gamma_{d_{C_q}}(\frac{\widetilde{\delta}+d_{C_q}-1}{2})}{\Gamma_{d_{C_q}}(\frac{\delta+d_{C_q}-1}{2})}\frac{\Gamma_{d_{S_q}}(\frac{\widetilde{\delta}+d_{S_q}-1}{2})}{\Gamma_{d_{S_q}}(\frac{\delta+d_{S_q}-1}{2})}
 \frac{\Gamma_{d_{C_{q_2}}}(\frac{\delta+d_{C_{q_2}}-1}{2})}{\Gamma_{d_{C_{q_2}}}(\frac{\widetilde{\delta}+d_{C_{q_2}}-1}{2})}\\
 &\times  \frac{\Gamma_{d_{C_{q_1}}}(\frac{\delta+d_{C_{q_1}}-1}{2})}{\Gamma_{d_{C_{q_1}}}(\frac{\widetilde{\delta}+d_{C_{q_1}}-1}{2})}.
\end{align*}
If using independent Bernoulli priors (with parameter $r$) for the edges included in $G$, $p(\widetilde{G})/p(G)=r/(1-r)$. The proposal ratio $p(\widetilde{G}\mid G)/p(\widetilde{G}\mid G)=(p(p-1)/2-n_e)/(n_e+1)$, with $n_e$ the number of edges in $G$. The likelihood ratio for the case of deleting is simply the inverse of that for the case of adding. 

With probability $p$, propose  $\widetilde{G} \sim \mbox{Unif}$, a (discrete) uniform distribution supported on the set of all decomposable graphs, and accept the proposal with probability
\[\alpha=\min\left\{1, \frac{p(\widetilde{G}\mid\{{\mc}_i^M\},\mc_0^M)}{p(G\mid\{{\mc}_i^M\},\mc_0^M)}\right\}.\]
\end{itemize}
Repeat step 1 for a large number of iterations until convergence is achieved.

\section*{2. More details on setting model parameters}
Several parameters need to be determined before applying Algorithm 1 or 2. The truncation parameters $\{M_j\}$ can be determined using some approximation criteria as discussed in the paper. The degrees of freedom $\delta$ of the $\hiwp_G$ prior of ${\bf Q}_{\mathcal{C}}$ is chosen as a positive integer. Smaller values of $\delta$ imply larger variances so that the prior is more ``vague.'' For the scale matrix ${\bf U}$ of the $\hiwp_G$ prior, we determine its value by first decomposing ${\bf U}={\bf Z R Z}$, where ${\bf Z}=\mbox{diag}\{\boldsymbol{\tau}\}$ is the marginal standard deviation of the basis coefficients. If using FPC analysis, $\boldsymbol{\tau}$ can be taken as the square root of the eigenvalues. In other cases, we suggest to choose $\boldsymbol{\tau}$ to be proportional to the (marginal) sample standard deviation, from the empirical Bayes perspective. The pattern of ${\bf R}$ can be hard to determine. We set ${\bf R}={\bf I}$ in our simulations and real data application. Other priors, like the Hyper-inverse Wishart g-prior of \citeasnoun{Carvalho2009}, would also be good options. In Algorithm 2, one also needs to determine the noise variance $\boldsymbol{\Lambda}$, whose value would influence the identification of ${\bf Q}_{\mathcal{C}}$. In this work, we have assumed additive white noise. Any orthogonal basis transform of Gaussian white noise is still white noise. The variance of the white noise in the frequency domain equals the corresponding variance in the time domain up to a scale parameter, which is approximately $|T_j|/(|\boldsymbol{\mt}_j|-1)$, where $|T_j|$ is the length of $T_j$ and $|\boldsymbol{\mt}_j|$ is the number of grid points on $T_j$. Therefore, we can estimate the white noise variance by firstly applying a localized linear smoother to the function, and then computing the sample variances of the residuals. This variance can then be transformed to the frequency domain. If using FPC analysis, the PACE algorithm of \cite{Yao2005} can be directly applied to compute the noise variances and eigenbasis, even for sparse functional data. For the initial values $\{{\mc}_i^M\}$ in Algorithm 2, one can simply set ${\mc}_i^M={\md}_i$. If the data are centered in a pre-processing step, one can set ${\mc}_0^M$ to be the zero vector; otherwise, one can use the sample mean of the estimated basis coefficients.

\section*{3. Methods for improving mixing}
Even though the small-world sampler in the MCMC Algorithms 1 and 2 helps improve mixing, as the number of vertices $p$ and the truncation parameters $\{M_j\}_{j=1}^p$ increase, the Metropolis-Hastings step may suffer low acceptance rate, causing slow convergence. More advanced Monte Carlo strategies, such as parallel tempering \cite{Liu2008}, may be adopted to further improve mixing.
Another alternative is the Small-world MCMC with Tempering algorithm proposed by Guan and Stephens ({http://arxiv.org/abs/1211.4675}), in which the heavy tailed proposal in the small-world sampler is replaced by a tempered version of the posterior distribution. 

\section*{4. More results for simulation 2}

\begin{figure}[h]
\centering
\includegraphics[width=5 in,angle=0]{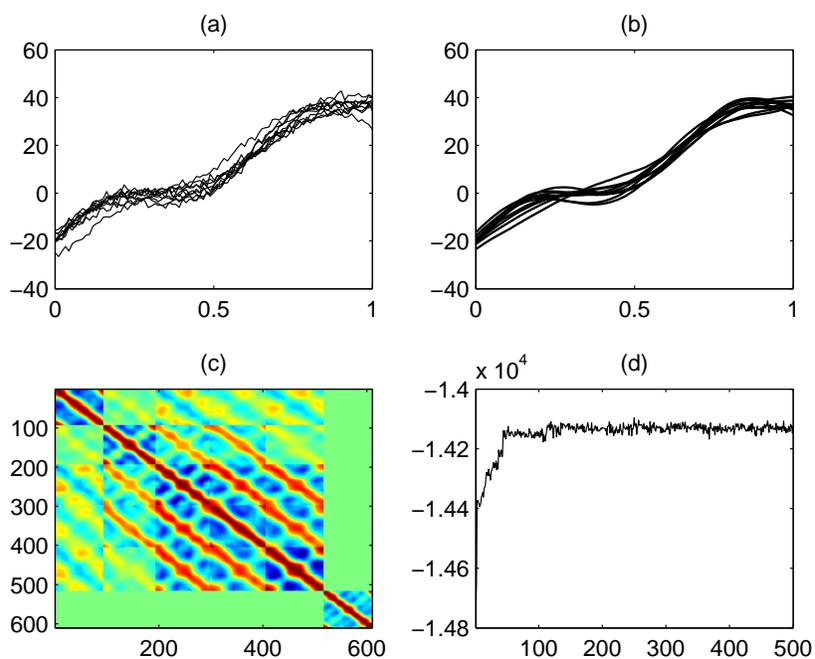}
\caption[Plots of Simulation 2]{Results for Simulation 2. (a): The plot of raw data for the first 10 samples of functional component 1. (b): The posterior mean estimate of $f_{i1}(t)$ corresponding to the curves in (a). (c): the posterior mean estimate of the data domain correlation matrix. (d): The trace plot of the log posterior densities of the first 500 samples.}
\label{fig:simu2}
\end{figure}

A plot of the noisy data is shown in panel (a) of Figure \ref{fig:simu2}, with its smooth estimates shown in Panel (b). The posterior estimate of the data domain correlation is plotted in panel (c), which corresponds to the true correlation plotted in (c) of Figure 2 in the main text. The trace plot of the conditional log posterior densities of the graph is shown in panel (d). 


\end{document}